\documentclass[reprint,
showpacs,
 amsmath,amssymb, aps, 
prb,
]{revtex4-1}

\pdfoutput=1 
\usepackage{pdfpages}
\usepackage{graphics}      
\usepackage{graphicx}      
\usepackage{dcolumn}  
\usepackage{slashed}
\usepackage{subfigure}
\usepackage{float}
\usepackage{epstopdf}
\usepackage{longtable}     
\usepackage{url}           
\usepackage{bm}            

\newcommand{\bra}[1]{\langle #1|}
\newcommand{\SWAP}{\text{SWAP}}
\newcommand{\I}{\text{I}}
\newcommand{\II}{\text{II}}
\newcommand{\ket}[1]{|#1\rangle}
\newcommand{\expect}[1]{\langle #1\rangle}
\newcommand{\plaqa}{
 {\mathchoice
 {\includegraphics[height=1.6ex]{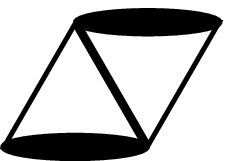}}
 {\includegraphics[height=1.6ex]{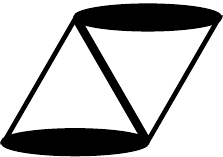}}
 {\includegraphics[height=1.2ex]{plaqa}}
 {\includegraphics[height=0.9ex]{plaqa}}
 }
}
\newcommand{\plaqb}{
 {\mathchoice
 {\includegraphics[height=1.6ex]{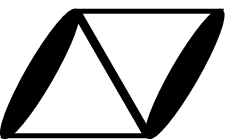}}
 {\includegraphics[height=1.6ex]{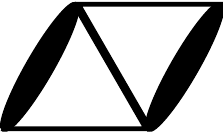}}
 {\includegraphics[height=1.2ex]{plaqb}}
 {\includegraphics[height=0.9ex]{plaqb}}
 }
}

\begin{document}

\title{Entanglement Entropy at Generalized Rokhsar-Kivelson Points of Quantum Dimer Models}

\author{Alexander Selem$^{1,2}$}
\email{aselem@berkeley.edu}
\author{C. M. Herdman$^{1,2,3}$}
\thanks{Present Address: Department of Physics, University of Vermont, VT 05405, USA}
\author{K. Birgitta Whaley$^{1,3}$}

\address{$^1$Berkeley Quantum Information \& Computation Center, University of California, Berkeley, CA 94720, USA}
\address{$^2$Department of Physics, University of California, Berkeley, CA 94720, USA}
\address{$^3$Department of Chemistry, University of California, Berkeley, CA 94720, USA}

\date{\today}

\begin{abstract}
We study the $n=2$  R\' enyi entanglement entropy of the triangular quantum dimer model via
 Monte Carlo sampling of Rokhsar-Kivelson(RK)-like ground state wavefunctions. Using the construction proposed by Kitaev and Preskill [Phys. Rev. Lett. 96, 110404 (2006)] and an adaptation of the Monte Carlo algorithm described by Hastings \emph{et al.} [Phys. Rev. Lett. 104, 157201 (2010)], we
compute the topological entanglement entropy (TEE) at the RK point $\gamma = (1.001 \pm 0.003) \ln 2$ confirming earlier results.
Additionally, we compute the TEE of the ground state of a generalized RK-like Hamiltonian and demonstrate the universality of TEE over a wide range of parameter values within a topologically ordered phase approaching a quantum phase transition. 
For systems sizes that are accessible numerically, we find that the quantization of TEE depends sensitively on correlations. We characterize corner contributions to the entanglement entropy and show that these are well described by shifts proportional to the number and types of corners in the bipartition.
\end{abstract}

\pacs{75.10.Jm, 05.30.Rt, 03.67.Mn}

\maketitle

\section{Introduction}
Quantum liquid phases of matter that do not break conventional symmetries can have ``hidden" non-local quantum orders. Such quantum liquids are ordered quantum phases that are not described by a local order parameter.  Topologically ordered phases in particular, are of great interest because of their potential to form the basis of a physically fault-tolerant quantum computer.~\cite{Nayak} There is therefore a strong incentive to realize such phases in experimental systems as well as to identify theoretical models which possess topologically ordered phases. 

However, the lack of a local order parameter inhibits the identification of topologically ordered phases in theoretical models. Kitaev and Preskill\cite{Kitaev} and Levin and Wen\cite{Levin} identified a sub-leading negative constant term in the bipartite entanglement entropy, the topological entanglement entropy (TEE), which allows for the identification and classification of topological order. Constant subleading terms can arise in other contexts including critical systems,~\cite{Hsu, Stephan2009, Stephan2011, Oshikawa} from Goldstone modes in symmetry-broken states,~\cite{Metlitski} and from corners in non-smooth bipartitions as seen in integer quantum Hall wavefunctions.~\cite{Rodriguez}  

Lattice models with hard local constraints, such as quantum dimer and loop models, possess quantum liquid ground state phases, including topological phases. In particular, the hard-core quantum dimer model on the triangular lattice (TQDM) has a $Z_2$ topologically ordered dimer liquid phase.~\cite{Moessner, Moessner_Sondhi} 
Since topological phases generally arise in strongly interacting systems which are not always tractable by analytic methods, numerical studies of these models are often necessary. Lanczos diagonalization may be used to compute the bipartite-entanglement entropy in small systems.~\cite{Furukawa} However, computing the sub-leading term in the entanglement requires using moderately large systems which are not generally accessible via Lanczos diagonalization. 

At the Rokhsar-Kivelson (RK) point the TEE of the TQDM has been computed using Pfaffian (Kasteleyn) methods with high ($10^{-9}$) numerical accuracy,~\cite{Stephan, Furukawa} and away from the RK point using exact diagonalization on small lattices.~\cite{Furukawa}
Recent work by Hastings \emph{et al}.~\cite{Hastings} has demonstrated a method for computing the R\' enyi entanglement entropy via Monte Carlo methods. This is attractive, since these techniques generally allow for the study of moderately large systems. 
 
Demonstrating the universality of the TEE within a topological phase and its behavior across phase transitions is an area of ongoing research. Temperature induced transitions have been explored in the work of Isakov \emph{et al}.~\cite{Isakov}
The behavior of TEE on approaching a quantum phase transition was previously studied by St\'{e}phan \emph{et al}. by interpolating between the triangular-and the square-lattice dimer models.~\cite{Stephan2011}

Here we adapt the method of Hastings \emph{el al}~\cite{Hastings} to the TQDM. We confirm the previous results for the TEE,~\cite{Stephan, Furukawa} and characterize constant contributions due to corner effects at the RK point which may compete with the TEE. Additionally we compute the TEE of a ``generalized" RK wave function, using the model of Trousselet \emph{et al}.,~\cite{Trousselet} and show the evolution of TEE approaching a first-order quantum phase transition. The results strongly suggest the universal nature of the TEE inside the dimer liquid phase in the thermodynamic limit, although correlations are found to limit convergence in finite systems.

\subsection{Triangular lattice quantum dimer model} 
\label{sec:dimer_intro}
The fully packed hard-core dimer model is defined on a lattice with degrees of freedom labeled by the occupation of dimers on links, and the constraint that exactly one dimer must touch each vertex.
The Hilbert space is comprised of the fully packed dimer coverings on the lattice
satisfying the vertex constraint ($\equiv \ket{\{C\}}$). Different dimer coverings are defined to be orthogonal. Rokhsar and Kivelson first introduced this model on the square lattice.~\cite{Rokhsar} It was subsequently generalized by Moessner and Sondhi~\cite{Moessner} to the triangular lattice, where indications of a $Z_2$ topologically ordered ground state emerge.~\cite{Moessner_Sondhi} The Hamiltonian for the TQDM is:
\begin{equation}
H= \sum\limits_{p}-t(\ket{\plaqa}\bra{\plaqb} +h.c.) +
 v(\ket{\plaqa}\bra{\plaqa}+\ket{\plaqb}\bra{\plaqb})
\label{eq:RK}
\end{equation}
where $p$ labels all minimal Rhombus plaquettes on the triangular lattice, and the kinetic ($t$) term which flips parallel dimers around a plaquette is the minimal dimer hopping that respects the vertex constraint (here $t>0$ always). The $v$ term acts as a potential energy between parallel dimers. At the RK point, defined by $v=t$, the ground state can be written as
\begin{equation}
\ket{\Omega} = \sum\limits_{C} \frac{1}{\sqrt{Z}} \ket{C}
\end{equation}
where the sum is over configurations reachable by plaquette flips ($Z$ is the number of elements in the sum). 
On the torus, plaquette flips conserve two parities that are defined by counting the occupation of dimers intersecting the two non-contractible loops of the torus. 
Dimer coverings are split into four topological sectors defined by these parities, such that local rearrangements of dimers cannot connect configurations in two different topological sectors. Plaquette flips on the triangular lattice are believed to be nearly ergodic within a topological sector, with the exception of 12 symmetry related ``staggered" configurations that have no flippable plaquettes. Therefore four distinct ground states are defined by these parities, which we label $\Omega= (0, 0)$, $(1, 0)$, $(0, 1)$, $(1, 1)$ (0 for even parity). This topological degeneracy is a characteristic of the topological order present in the ground state at the RK point.

On the triangular lattice the topologically ordered dimer liquid phase persists for a finite region below the RK point, in the range $0.86 \simeq v/t \leq 1$. \cite{Moessner, Ralko} Outside of this region ($v/t \lesssim 0.86$ and $v/t>1$), the ground state is one of several symmetry-broken ordered crystalline phases.\cite{Ralko}

The RK wave function can be generalized to weighted superpositions of dimer configurations $\{ C \}$, 
\begin{equation}
\ket{ \Psi}= \sum\limits_{C} \frac{1}{\sqrt{Z}} e^{-E(C)} \ket{C},
\end{equation}
where $Z=\sum\limits_{C} e^{-2E(C)}$, and $E(C)$ is the ``classical" energy of $C$. Such a generalized RK wave function is the exact zero-energy ground state of a corresponding RK-like local Hamiltonian.~\cite{Castelnovo} In Sec.~\ref{sec:TEE}B, we compute the TEE of a generalized RK wave function that was previously studied in Refs. \citenum{Castelnovo} and \citenum{Herdman}, and seen to interpolate between the topologically ordered phase and a symmetry-broken phase.

\subsection{Entanglement entropy and topological phases}
Bipartite entanglement entropy has emerged as a powerful probe of quantum systems.  
The bipartite entanglement entropy of a pure state $\ket{\Psi}$ is defined with respect to a bipartition of the lattice into a region $A$ and its complement $B$. The von Neumann entropy is defined as

\begin{equation}
S(\rho_A) \equiv -\text{Tr} \rho_A \ln \rho_A
\end{equation}
and the R\'{e}nyi entropy is defined as
\begin{equation}
S_{n}(\rho_A) \equiv \frac{1}{1-n} \text{Tr}\rho_A^n,
\end{equation}
where $\rho_A \equiv \text{Tr}_B |\Psi \rangle\langle\Psi |$ is the reduced density matrix of $A$. The R\'{e}nyi entropy reduces to the von Neumann entropy in the limit $n \rightarrow 1$ and both are symmetric under exchange of $A$ and $B$, $S_n(\rho_{A})=S_n(\rho_{B})$.
Ground states of local Hamiltonians are known to exhibit a boundary law scaling in region size [i.e., in two dimensions (2D) the scaling is with the perimeter length],\cite{Eisert} although critical fermions are a notable exception to this rule.~\cite{Wolf} In two dimensions this scaling can generically be written as  
\begin{equation}
S(\rho_A) = \alpha L_A + \beta\log(L_A/a) +C_0 + O(1/L_{A}),
\end{equation}
where the leading term is proportional to the perimeter $L_A$, and $\alpha$ is a non-universal constant. The logarithmic term appears in certain quantum critical theories (the base is unspecified since differences can be absorbed into $\beta$). However for gapped phases, it is expected that $\beta=0$. The constant term $C_0$ has been shown to arise in critical phases as well as in topologically ordered phases.~\cite{Levin, Kitaev, Hsu, Stephan2009, Stephan2011, Oshikawa}

For topological phases, there is a universal, negative, constant subleading term, the topological entanglement entropy: $-\gamma\in C_0$ (with $\gamma$, also referred to as $\gamma_{topo}$, $>0$).\cite{Levin, Kitaev} Topological phases may be described by an effective topological quantum field theory;~\cite{ Levin2005, Nayak} such theories are categorized by the so-called total quantum dimension $D$. For conventional ordered phases $D=1$ and for topologically ordered phases $D>1$. The TEE is given by:
\begin{equation}
\gamma =\ln D
 \end{equation}
and therefore is a witness of topological order ($\gamma = 0$ for conventional phases).

Physically the origin of this term can be seen by considering string-net wavefunctions as an effective theory of topological order.~\cite{Levin2005} The non-local order encoded in a topologically ordered phase can be understood in terms of effective loop or string-net degrees of freedom describing the wave function. Specifically, for discrete gauge theories, wavefunctions are comprised of different types of non-branching loops with (counting the absence of a loop as one type) the relation: \emph{the number of types of loops is equal to the elements of the group which is equal to the total quantum dimension}, $D$. Then as a direct consequence of the fact that each type of loop must enter and exit the boundary an even number of times, the effective degrees of freedom crossing the bipartition boundary as probed by the entanglement entropy, is corrected by a factor of $1/D$. This is responsible for the reduction of the entanglement entropy scaling by $\ln D$.

The dimer model belongs to the $Z_2$ class~\cite{Moessner_Sondhi} of topologically ordered phases~\cite{Hamma, Levin, Kitaev} (the effective loop degrees of freedom are so-called transition loops;~\cite{Sutherland, Kohmoto, Herdman} see the Supplemental Material~\cite{SM}). Therefore for the dimer model and other $Z_2$ topologically ordered phases $\gamma = \ln 2$.  Furthermore it has been shown~\cite{Flammia} that $\gamma$ is independent of the R\'{e}nyi parameter $n$, so that any R\'enyi entanglement entropy can be used to compute the quantity $\gamma$. 

In Sec.~\ref{sec:corners} we show that for non-smooth bipartitions on a lattice, there can be non-universal constant contributions to $C_0$. We will split $C_0$ into universal and non-universal parts by writing $C_0= -\gamma + \kappa$. Our results are consistent with a non-universal term $\kappa$ of the form $\kappa = \sum a_i n_i$, where $n_i$ is the number of corners of type $i$. As discussed in Sec.~\ref{sec:corners}, these corner terms can be thought of as coming from a substitution in a generalized linear scaling $S_A \sim \sum_i \alpha_i \ell_i$, where the boundary vertices $i$ contribute different constants $\alpha_i$.

\section{Monte Carlo Sampling for Entanglement Entropy}
\label{sec:methods}
In Ref.~\onlinecite{Hastings} Hastings \emph{et al}. describe a SWAP algorithm to compute $S_2(\rho_A)$ via Monte Carlo simulations [in the following, $S_A$ will always be taken to mean $S_2(\rho_A)$]. 
In the current work we will be considering generalized RK points, characterized by wave functions that are explicitly written as a weighting of configurations. Therefore we are able to compute expectation values of estimators using classical Monte Carlo sampling of the wave function. 

To estimate the entanglement entropy, following Ref.~\onlinecite{Hastings}, we define a new ``doubled'' system as two non-interacting independent copies of the original, labeled 1 and 2. Each corresponding copy has the identical bipartition $A$ and $B$ so that the Hilbert space of the doubled system is a tensor product of the two copies with a state labeled by degrees of freedom in $A_1$, $B_1$ and $A_{2}$, $B_{2}$, respectively. Then $S_A$ is related to the expectation of the $\SWAP_A$ operator defined on the doubled Hilbert space by its action in swapping the degrees of freedom in A: 
\begin{equation}
\SWAP_A \ket{A_1 B_1} \otimes \ket{A_2 B_2} = \ket{A_2 B_1} \otimes \ket{A_1 B_2}.
\end{equation}
Hastings \emph{et al}. showed that $\text{Tr}\rho_A^2 = \expect{\SWAP_A}$, and therefore $S_A = - \ln \expect{ \SWAP_A}$. 

Taking $C$ to represent a ``doubled'' dimer covering, the matrix elements of the SWAP operator are
\begin{equation}
\langle C' | \SWAP_A | C \rangle = \delta_{C'C_A} \delta(C|_A),
\end{equation} 
where $C_A$ is the configuration resulting from swapping $C$ over region A, and $\delta(C|_A)$ is 1 if the swapped configurations do not violate the hard-core dimer constraint, and zero otherwise. We can write the expectation value of SWAP as a weighted sum over configurations $C$:
\begin{equation}
\begin{split}
\expect{ \SWAP_A } &= \frac{1}{Z}\sum\limits_{C', C} e^{-E(C')-E(C)} \langle C' | \SWAP_A | C \rangle \\
&= \frac{1}{Z}\sum\limits_{C} e^{-E(C_A)-E(C)}\delta(C|_A) \\
&= \sum\limits_{C} e^{-\Delta E(C\vert_A)}\delta(C|_A) \Pi \left( C \right),
\end{split}
\label{eq:swap}
\end{equation}
where $Z=\sum_C \exp[-2E(C)]$, $\Pi \left( C \right) \equiv \exp[-2E(C)]/Z$ can be viewed as a probability distribution, and $\Delta E(C\vert_A) \equiv E(C_A)-E(C)$.
 We see then that by classical Monte Carlo sampling of $\Pi \left( C \right)$, we can compute the expectation value of the SWAP operator with use of the estimator $e^{-\Delta E(C)}\delta(C|_A)$. For the RK ground state, the expectation value of the swap operator is simply the fraction of the dimer configurations that are $A$-swappable.

As a direct consequence of the perimeter law scaling, the principal limitation of this approach is the exponential decay of $\expect{\SWAP_A}$ with boundary length. Following Ref.~\onlinecite{Hastings}, a ``ratio method" may be employed in which the ratio of expectation values of SWAP can be computed more efficiently:
\begin{equation}
 \frac{\expect{\SWAP_{A'}}}{\expect{\SWAP_{A}}}=e^{(S_{A}-S_{A'})} \sim e^{-\alpha(L_{A'}-L_A)}, 
\label{eq:ratio1}
\end{equation}
where region $A'$ is larger than $A$ but with perimeter $L_{A'}$ sufficiently close to $L_A$ such that the ratio is not too small. While a single ratio combination gives the \emph{difference} of the entanglement entropy of two regions, the entanglement entropy of a single (large) region, required to compute the TEE, can be determined by computing the entropy of a single small region and adding successively the computation of differences (swap ratios): [i.e. $S_{An} = S_{A0}+\sum_{i=1}^n (S_{Ai}-S_{A(i-1)})$, for $A_n > \dots > A_i >\dots > A_0$].

\begin{figure} 
\includegraphics[width=.27\textwidth]{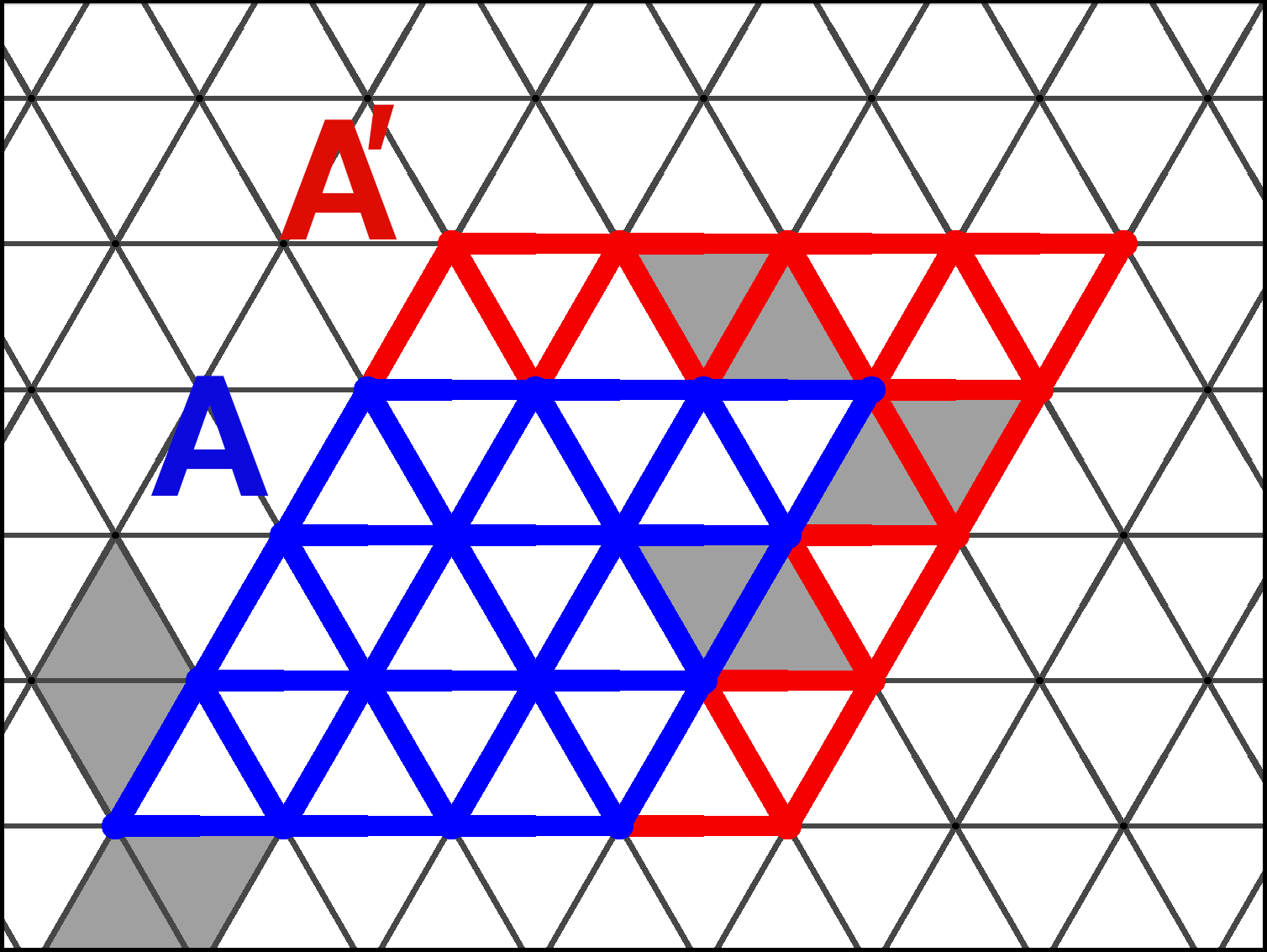} 
\caption{(Color online) Regions $A'$ and $A$ as described in the text. Note that $A'$ also includes the blue links. Shaded plaquettes are examples of constrained plaquettes for the case where $A$ is kept flippable.}
\label{fig:ratio}
\end{figure}

The ratio method can be specialized to the dimer system at generalized RK points. An example of two overlapping regions $A'$ and $A$ is shown in Fig.~\ref{fig:ratio}. 
Because the estimator used to compute $\expect{\SWAP_A}$ [Eq.~(\ref{eq:swap})] does not have support everywhere in the doubled Hilbert space
(it is non-zero only for swappable configurations), a difficulty with directly computing the ratio in Eq.~(\ref{eq:ratio1}) is that swappable configurations over $A'$ \emph{are not a subset} of those over $A$ and vice versa.
Instead, what is actually directly possible to compute are the following ratios: 
\begin{equation}
R'=\frac{\expect{ \SWAP_{A'} \SWAP_A} }{\expect{ \SWAP_{A'} }}, \:\; R=\frac{\expect{  \SWAP_{A'} \SWAP_A} }{\expect{  \SWAP_{A} }}. 
\end{equation}
From these, the ratio we want is simply $R/R'$. Each of these ratios has a simple Monte Carlo interpretation. For example, inserting $1= \sum_C \ket{C}\bra{C}$ into the numerator of $R$, and acting each $\SWAP$ towards the center readily leads to
\begin{equation}
\begin{split}
R=&\frac{\sum\limits_{C} \delta(C|_{A}) \delta(C|_{A'})e^{(-E(C_{A})-E(C_{A'}))}}
{ \sum\limits_{C} e^{-E(C_A)-E(C)}\delta(C|_A)  } \\
=&\sum\limits_{C|_{A}}    e^{-\Delta E(C\vert_{A'}) } \delta(C|_{A'}) \Pi_A(C)
\end{split}
\end{equation} 
where the sums have been explicitly written as restricted to $A$-swappable configurations, $\Delta E(C\vert_{A'})\equiv E(C_{A'})-E(C)$, and
\begin{equation}
\Pi_A(C)= \frac{\exp(-E(C_A)-E(C))}{\sum\limits_{C|_{A}}\exp(-E(C_A)-E(C))}.
\end{equation} 
To compute $R$, we sample dimer configurations (weighted by $e^{-E(C_A)-E(C)}$) in such a way that the $A$-boundary always remains swappable, and then measure the estimator $\delta(C|_{A'})  e^{-\Delta E(C\vert_{A'}) }$. At the standard RK point one may set $\exp \rightarrow 1$, and $R$ is simply the ratio of configurations that are swappable for both $A$ and $A'$ to those only $A$-swappable. 
To compute $R'$, the region which is kept swappable is reversed.

To generate configurations which always remain swappable over $A$,  updates are done independently for each copy, except for certain constrained plaquettes which must be flipped simultaneously in copies 1 and 2. The constrained plaquettes are those which are not entirely in region $A$ or $B$ (the complement of $A$). Some examples are shown in Fig.~\ref{fig:ratio}. For simply-connected bipartitions these updates should be ergodic over all possible $A$-swappable configurations. However, if $A$ is not simply connected, it can be shown that this is not the case, and another method for ratio updates is needed. Details are presented in the Supplementary Material.~\cite{SM} 

\section{Computation of Topological Entanglement Entropy}
\label{sec:TEE}

The TEE has been computed at the RK point in Ref.~\onlinecite{Stephan, Furukawa} using Kasteleyn matrices.~\cite{Kasteleyn} Here we report computations of this quantity using Monte Carlo simulation, which also allows us to sample generalized RK points whose ground states exhibit a quantum phase transition. Our findings are useful for further numerical studies away from RK-like wavefunctions.  

There are two routes to compute the TEE term numerically. The first is to use extrapolations: that is, to extrapolate the linear part of the entanglement entropy scaling for different perimeter-sized bipartitions and deduce the intercept. The second is to make use of cancellations: i.e., to consider a difference of bipartitions whose total net perimeter cancels while the net \emph{number} of boundaries does not, thereby leaving the topological contribution $\gamma$. 

\begin{figure} 
\includegraphics[width=.17\textwidth]{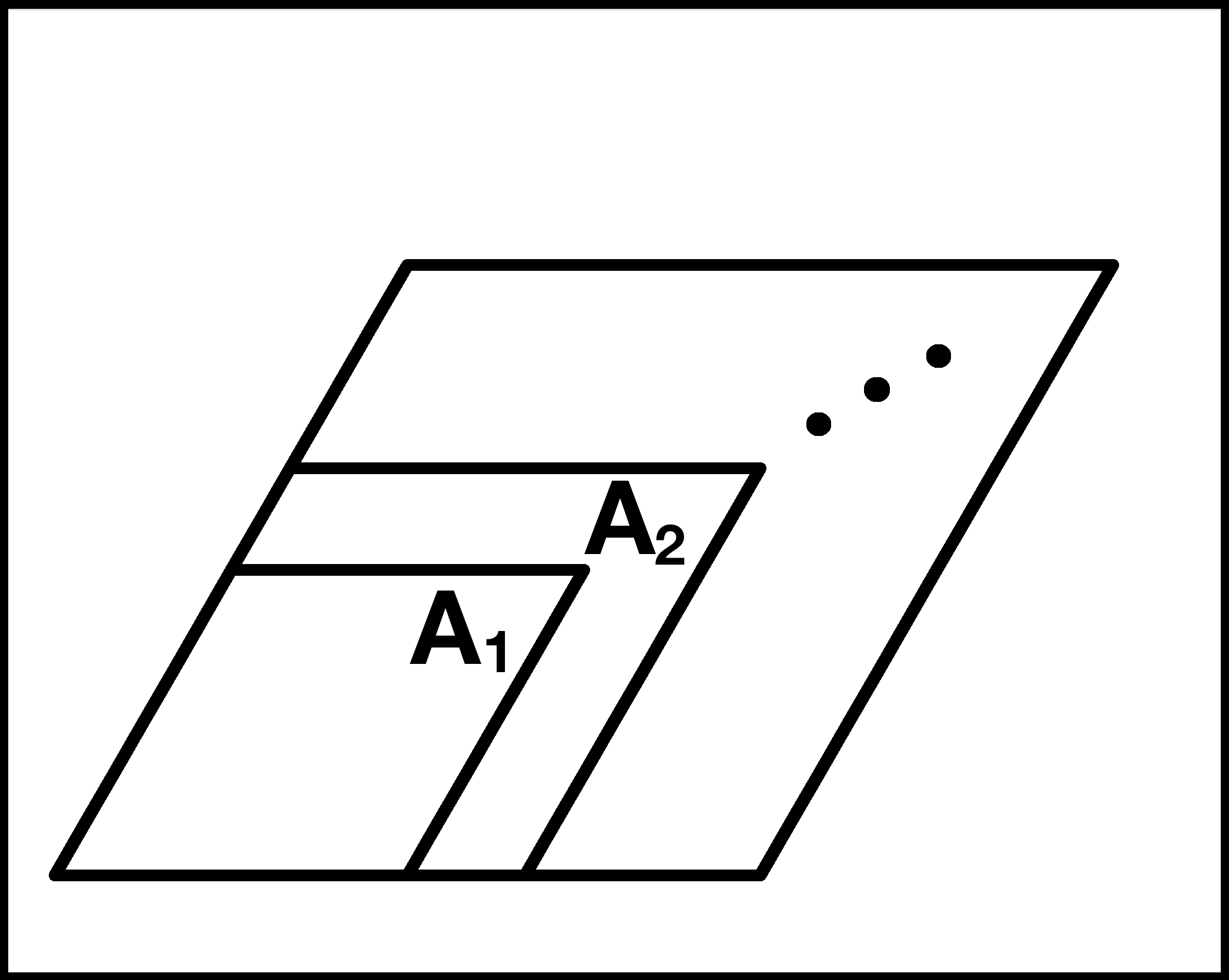} 
\; \; \;
\includegraphics[width=.17\textwidth]{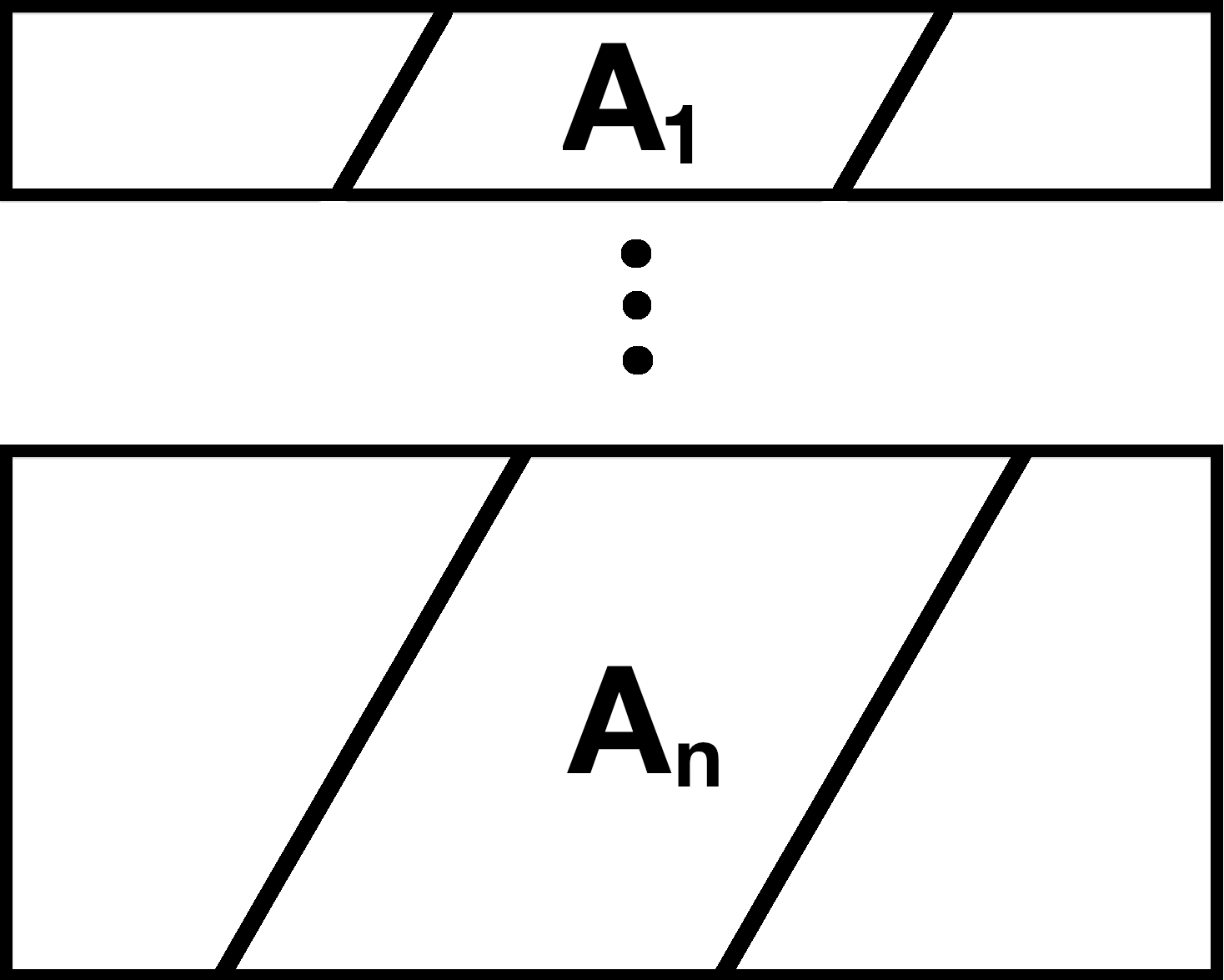} 
\caption{ Sample parallelogram extrapolation (left); strips with periodic boundary conditions assumed (right).}
\label{fig:extrap}
\end{figure}

When using extrapolations, one has the further option of employing simply connected polygonal bipartitions such as parallelograms of various sizes, or splitting the torus---the topology of periodic boundary conditions---into two pieces with two separate smooth boundaries. In the latter, each region is a non-simply-connected strip and tori of different widths must be used for boundary scaling (Fig.~\ref{fig:extrap}). Polygonal bipartitions suffer from corner contributions masking the TEE, which we describe in detail in Sec.~\ref{sec:corners}. Extrapolations from strips, which have smooth boundaries, avoid these corner effects. Also these types of non-trivial bipartitions can yield more information regarding the topological phase, such as $S$-matrix terms.~\cite{Zhang, Dong, Fradkin2009}  In recent work,\cite{Balents} a single smooth bipartition of a \emph{cylinder} has been employed to detect TEE using the density matrix renormalization group (DMRG). However, there are difficulties with this approach using current methods. First, we found that strips on thin tori appear to have substantial even-odd finite-size effects. Also, it is more difficult to formulate an appropriate ratio method for this strip geometry, given the ergodicity problems discussed in the Supplemental Material.~\cite{SM} Therefore, the alternative cancellation strategy turns out to be more useful for Monte Carlo simulations. 

\begin{figure} 
\includegraphics[width=.18\textwidth]{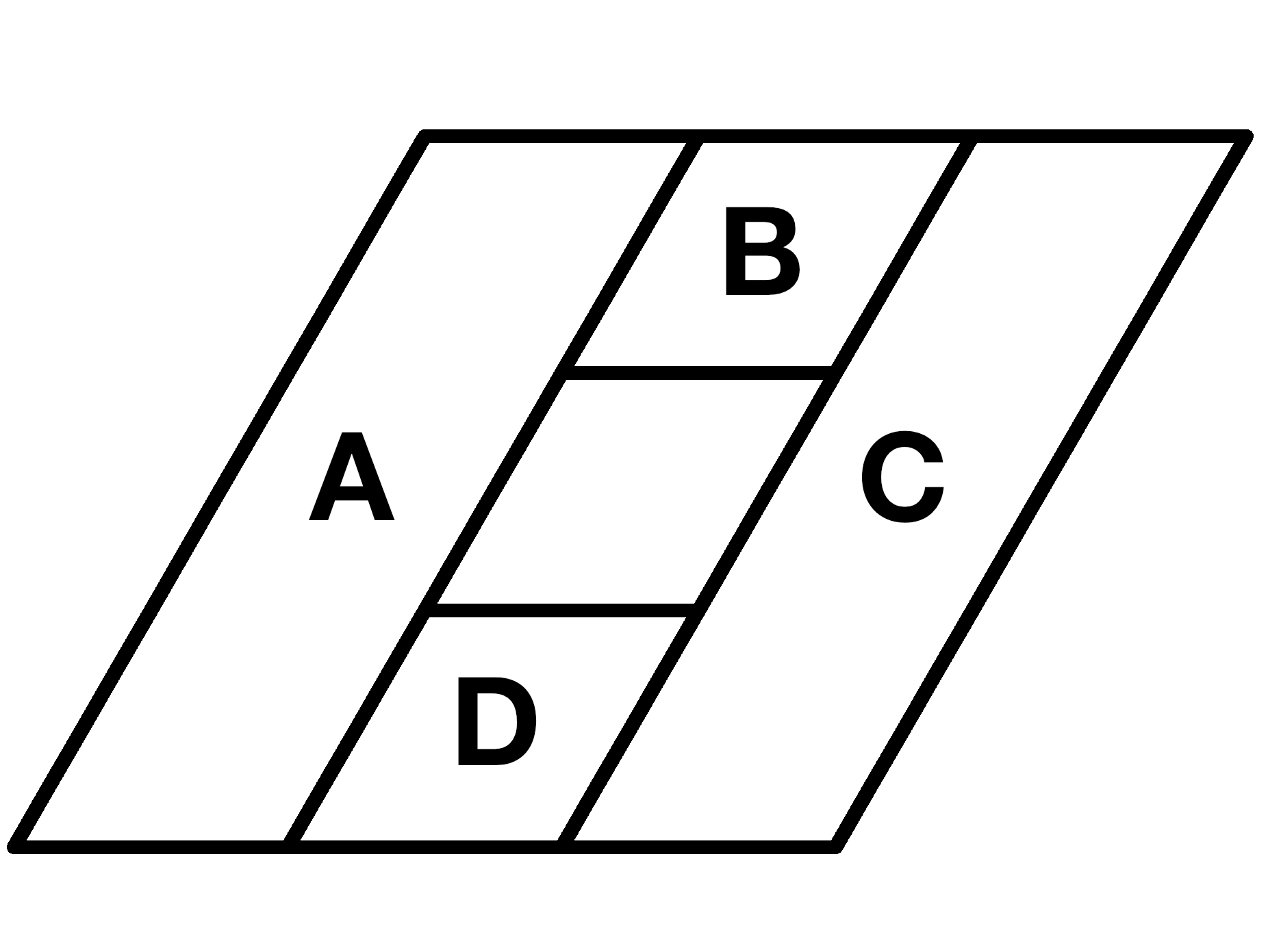} 
\includegraphics[width=.19\textwidth]{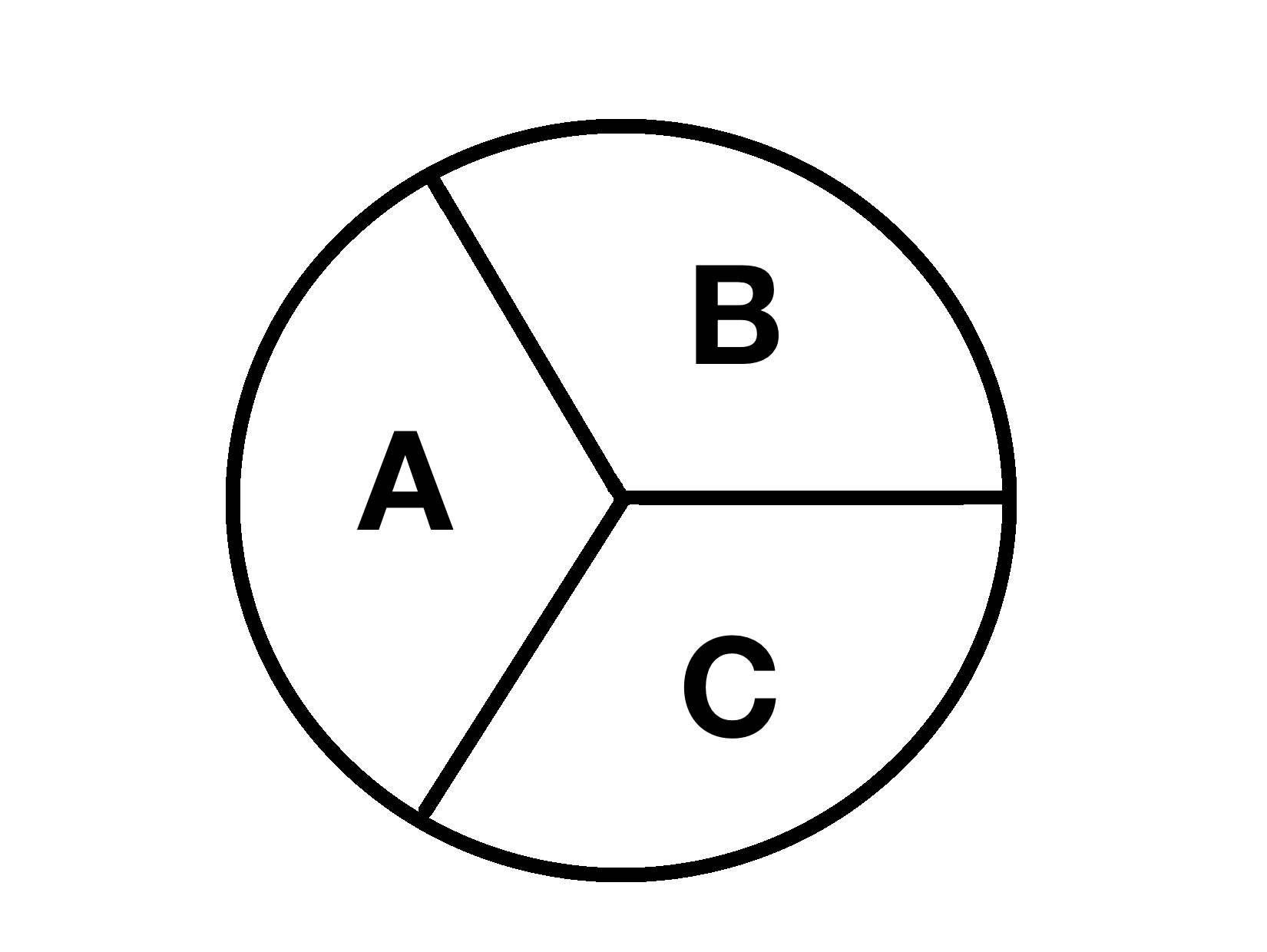} 
\caption{Levin-Wen (left) and Kitaev-Preskill (right) constructions.}
\label{fig:cancellation}
\end{figure}

Two cancellation geometries are typically used: a Levin-Wen-type construction~\cite{Levin} shown in Fig.~\ref{fig:cancellation} gives the TEE as two pairs $-2\gamma = (S_{ABCD} -S_{ABC}) + (S_{AC} -S_{ADC})$, while the Kitaev-Preskill construction~\cite{Kitaev} (Fig.~\ref{fig:cancellation}) gives the TEE as three pairs plus an extra region: $-\gamma = (S_{ABC} -S_{AB}) + (S_{A} -S_{AC})+ (S_{B} -S_{BC})+S_C$.  It turns out that in both of these constructions the number and type of corners cancel as well. 


In practice, however, using the types of updates described in Sec.~\ref{sec:methods} the Levin-Wen construction turns out to not be possible for the ratio method. The issue is that the updates fail to be ergodic in the configuration space. The problem occurs for calculations when the regions $S_{ABCD}$ and $S_{AC}$ are the constrained regions (the denominators in the ratios of Sec.~\ref{sec:methods}; see the Supplemental Material~\cite{SM}). Consequently the Kitaev-Preskill construction offers the best method for extraction of $\gamma$. 

\subsection{RK point}
\begin{figure} 
\includegraphics[width=.3\textwidth]{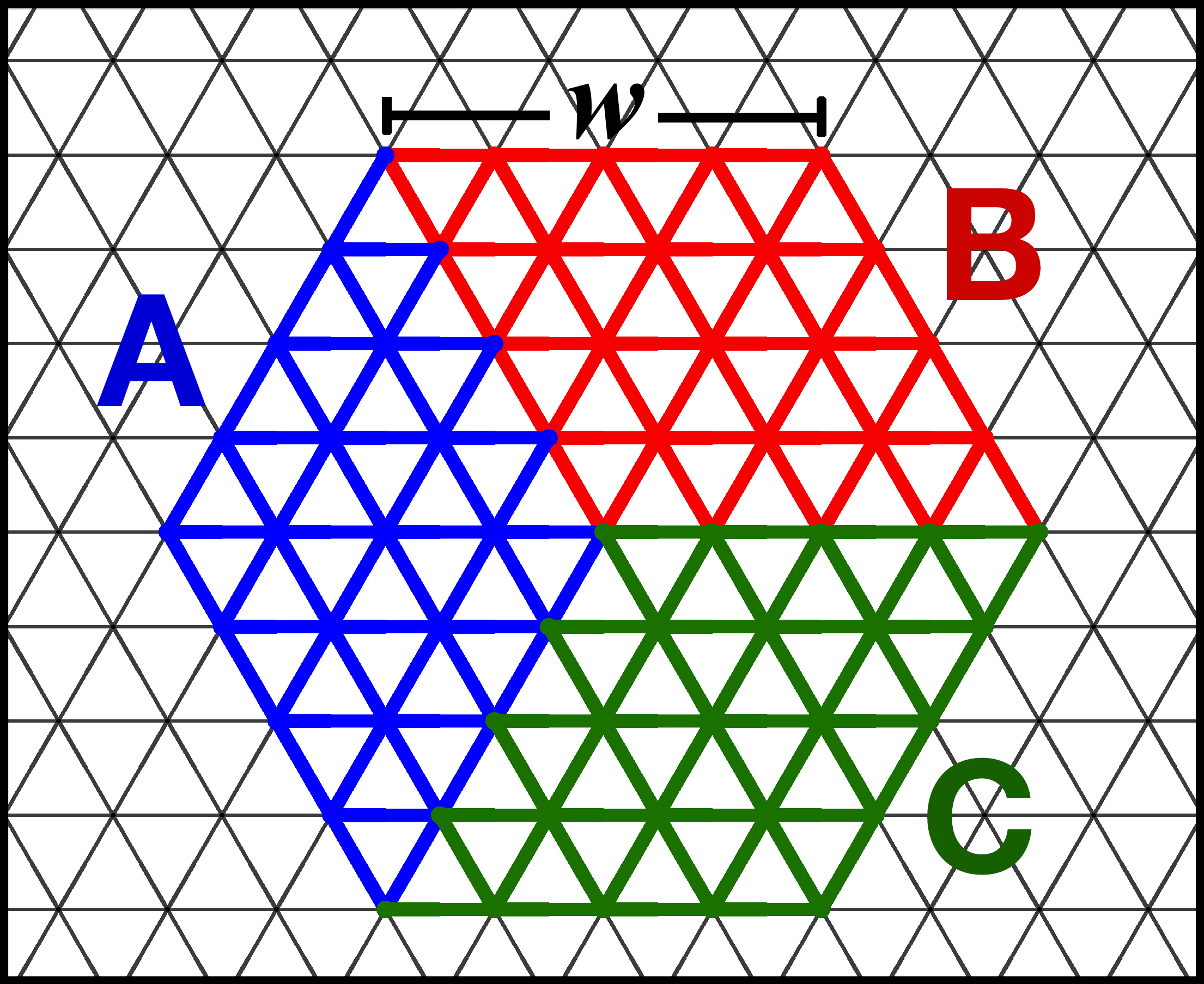} 
\caption{(Color online) Triangular lattice version of the Kitaev-Preskill construction used in the Monte Carlo runs.}
\label{fig:KP}
\end{figure}

\begin{figure} 
\includegraphics[width=.45\textwidth]{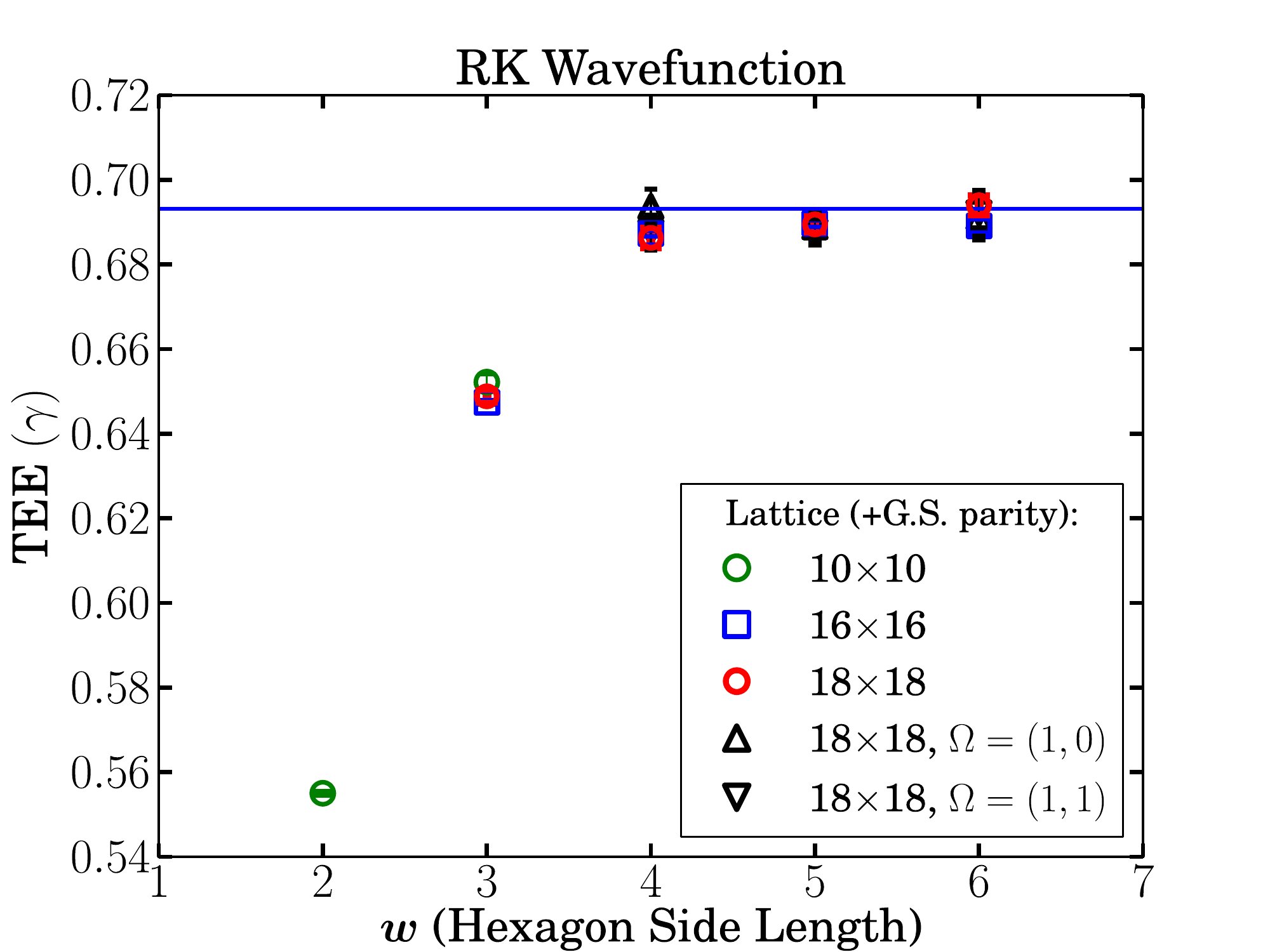}
\caption{(Color online) Net entanglement entropy (TEE) from the Kitaev-Preskill construction, for the RK wavefunction on the triangular lattice, computed using the partitions shown in Fig.~\ref{fig:KP}. Unless noted, the ground-state parity $\Omega$ is $(0, 0)$. The solid horizontal line is the predicted value of the TEE ($= \ln2$).}
\label{fig:RK}
\end{figure}

We take region $ABC$ to be an inscribed regular hexagon with divisions as shown in Fig.~\ref{fig:KP}. Note that since the Hilbert space is defined on links, partitions are uniquely defined by specifying to which region every link belongs. $w$ is the outer hexagon's side width. Our results for the RK point wavefunction are summarized in Fig.~\ref{fig:RK} and Table~\ref{tab:RK} for various sized hexagons and lattices.  If the spacing from the outer hexagon edge to the lattice edge is kept on the order of the hexagon side width or greater, the resulting TEE depends little on lattice sizes, except in limiting the size of the outer hexagon that can be used (this turns out not to be the case for more general wavefunctions). For $w=6$ convergence to the expected value of $\ln 2$ is seen (within the error bars). We also computed the TEE for the different ground states on the torus, $\Omega= (1, 0), (1, 1)$ (in the notation of Sec.~\ref{sec:dimer_intro}). These states also converge to the expected value, $\ln 2$. 

\begin{table} 
\caption{Numerical values for selected points of Fig.~\ref{fig:RK}.}
\label{tab:RK}
\newcolumntype{.}{D{.}{.}{12}}
\begin{tabular}{|ccc.|}
\hline
Monte Carlo run& \multicolumn{3}{|c|}{ $\gamma/\ln 2$}\\
\hline
$w=6$ ($18 \times 18$ lattice)  && &1.001\pm  0.003 \\
$w=6$ ($16 \times 16$ lattice)  && &0.994 \pm0.002   \\
$w=5$ ($18 \times 18$ lattice)   && &0.995\pm 0.003 \\  
$w=5$ ($16 \times 16$ lattice)   && &0.995\pm0.001  \\    
\hline
\end{tabular}
\end{table} 

\subsection{TEE approaching a quantum phase transition: Generalized RK wavefunction}

A number of phases from other classical weightings have been studied.~\cite{Trousselet, Castelnovo, Herdman} One class of modifications interpolates between the liquid-like RK point and symmetry-broken phases by preferentially weighting dimers on links reflecting the desired ordered configuration.~\cite{Herdman} These wavefunctions, however, explicitly break translation invariance and therefore constructions which rely on perimeter cancellations fail to accurately probe the TEE unless very large regions can be used. 

A classical dimer model that preserves translation invariance was discussed by Trousselet \emph{et al}. in Ref.~\onlinecite{Trousselet}. In this model the wavefunction was weighted to favor dimers in parallel (equivalent to a flippable plaquette), as if they would ``interact classically'' in the sense of $E(C)$ as a classical energy. In this case, $E(C)=\ln({\alpha}) N_{f}(C)$, where $N_{f}(C)$ is the number of flippable plaquettes for a covering $C$, and $\alpha$ is an adjustable parameter. $\alpha=1$ corresponds to the RK point while $\alpha<1$ favors flippable plaquettes. Such interacting wavefunctions are the groundstate of the RK-like Hamiltonian\cite{Castelnovo,Herdman} 
\begin{equation}
\begin{split}
H_{\alpha}= t\sum\limits_{p}&-(\ket{\plaqa}\bra{\plaqb} +H.c.) \;\; \\
 +&(\alpha^{-\Delta N_f/2}\ket{\plaqa}\bra{\plaqa}+\alpha^{\Delta N_f/2}\ket{\plaqb}\bra{\plaqb}) 
\label{eq:inter}
\end{split}
\end{equation}
with $\Delta N_f = N_f(\plaqb)-N_f(\plaqa)$ (the difference in flippable plaquettes due to a flip of plaquette $p$). 

The topological liquid phase with a finite correlation length persists for $\alpha<1$ until the wavefunction undergoes a first order phase transition near $\alpha_c \sim 0.2$.\cite{Herdman, Trousselet}  Below this value only configurations that have the maximum number of flippable plaquettes contribute to the wavefunction. These are $12$ columnar symmetry-broken configurations, plus a large number of configurations related by single line shifts across the lattice.~\cite{Moessner, Trousselet} 

Using the ratio method and the Kitaev-Preskill construction, we are able to probe the TEE of such interacting wavefunctions in the liquid regime as the system approaches the transition point. Although the TEE is believed to be universal in the topological liquid regime, there is to date only one known example at $T=0$, namely, that given by St\'{e}phan \emph{et al}. in Ref.~\onlinecite{Stephan2011}. These authors computed the TEE for a dimer model starting on the triangular lattice (at the RK point) and interpolated to the square lattice using Kasteleyn matrices.~\cite{Kasteleyn, Fisher} At a critical point the entanglement entropy is predicted to have a constant positive shift. However Stephan \emph{et al}. found that the TEE decreased below $-\ln 2$ before rising toward a positive value. The flow was well described by a single combination of parameters $t L$, where $L$ is simultaneously the cylinder circumference (the geometry used) and bipartition boundary, while $t$ is the fugacity for dimers on diagonal bonds ($t=0$ is the square lattice fugacity; note that this is a different $t$ from the one in Eq.~(\ref{eq:RK})]. Deviation away from $\ln 2$ begins for as large a value as $t L \sim 9$ [for R\'{e}nyi parameter $n=1.5$).

\begin{figure} 
\includegraphics[width=.45\textwidth]{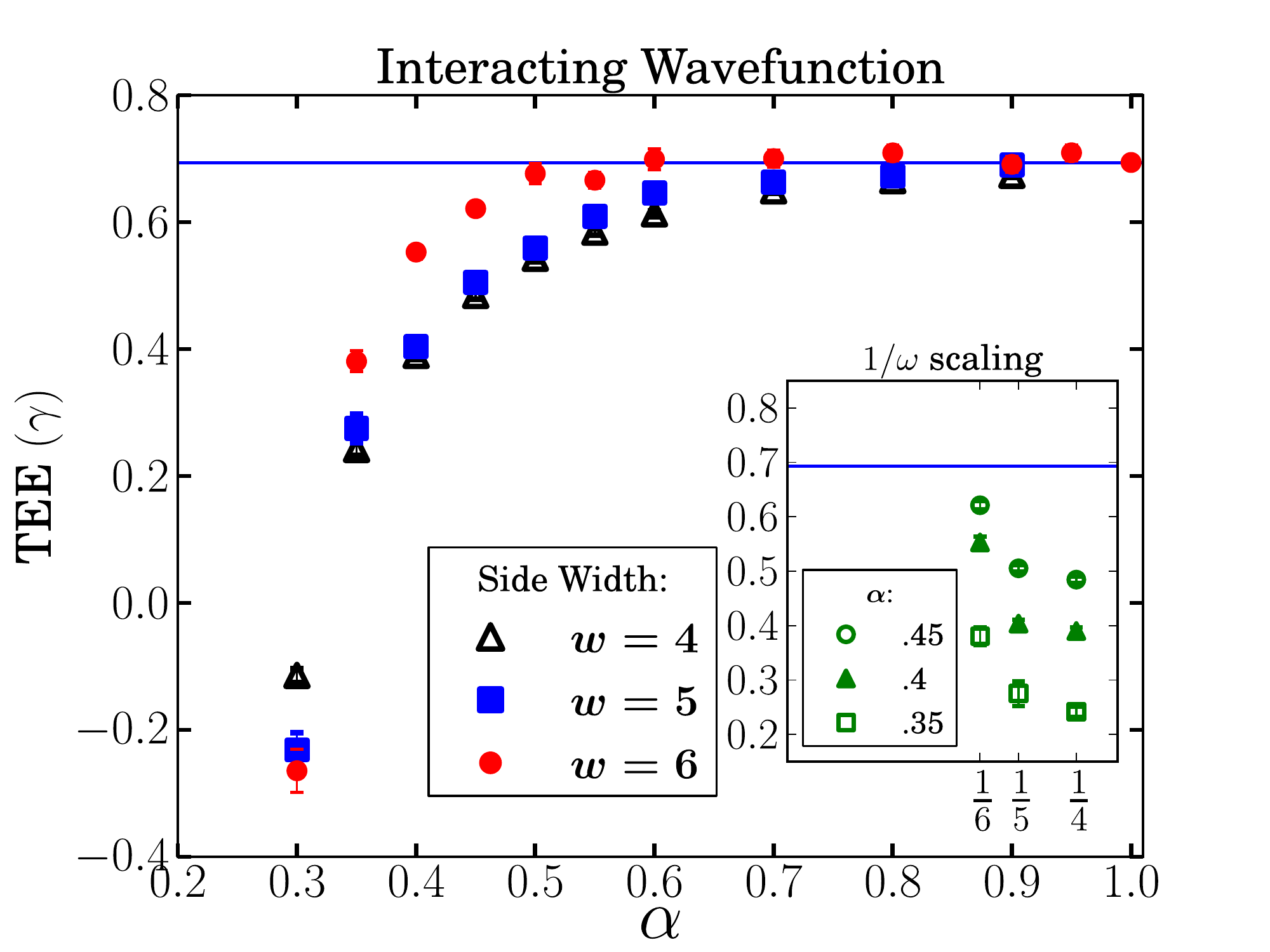}
\caption{(Color online) TEE for the ground state of Eq.~\ref{eq:inter} as a function of the parameter $\alpha$. All points are computed on an $18\times18$ lattice, with the TEE extracted from a Kitaev-Preskill construction with outer hexagon width $w$.}
\label{fig:NFPLQ}
\end{figure}

Our results for the interacting wave function are presented in Fig.~\ref{fig:NFPLQ} as a function of the parameter $\alpha$. Since the ordered state involves many local minima, we cannot go through the transition without non-local updates, which we have not implemented in the current variant of the ratio method.  The TEE appears to be a robust indicator of topological order for a reasonable range of $\alpha$. However, below $\alpha\sim 0.5$ stronger finite size effects begin to affect the result. An investigation of dimer-dimer correlations, discussed below, suggests that larger correlations are the main reason for the deviation. While the scaling with outer hexagon width $w$ is not clear for the system sizes studied (inset to Fig.~\ref{fig:NFPLQ}), it suggests convergence to the expected value of $\ln 2$.
\begin{figure} 
\includegraphics[width=.42\textwidth]{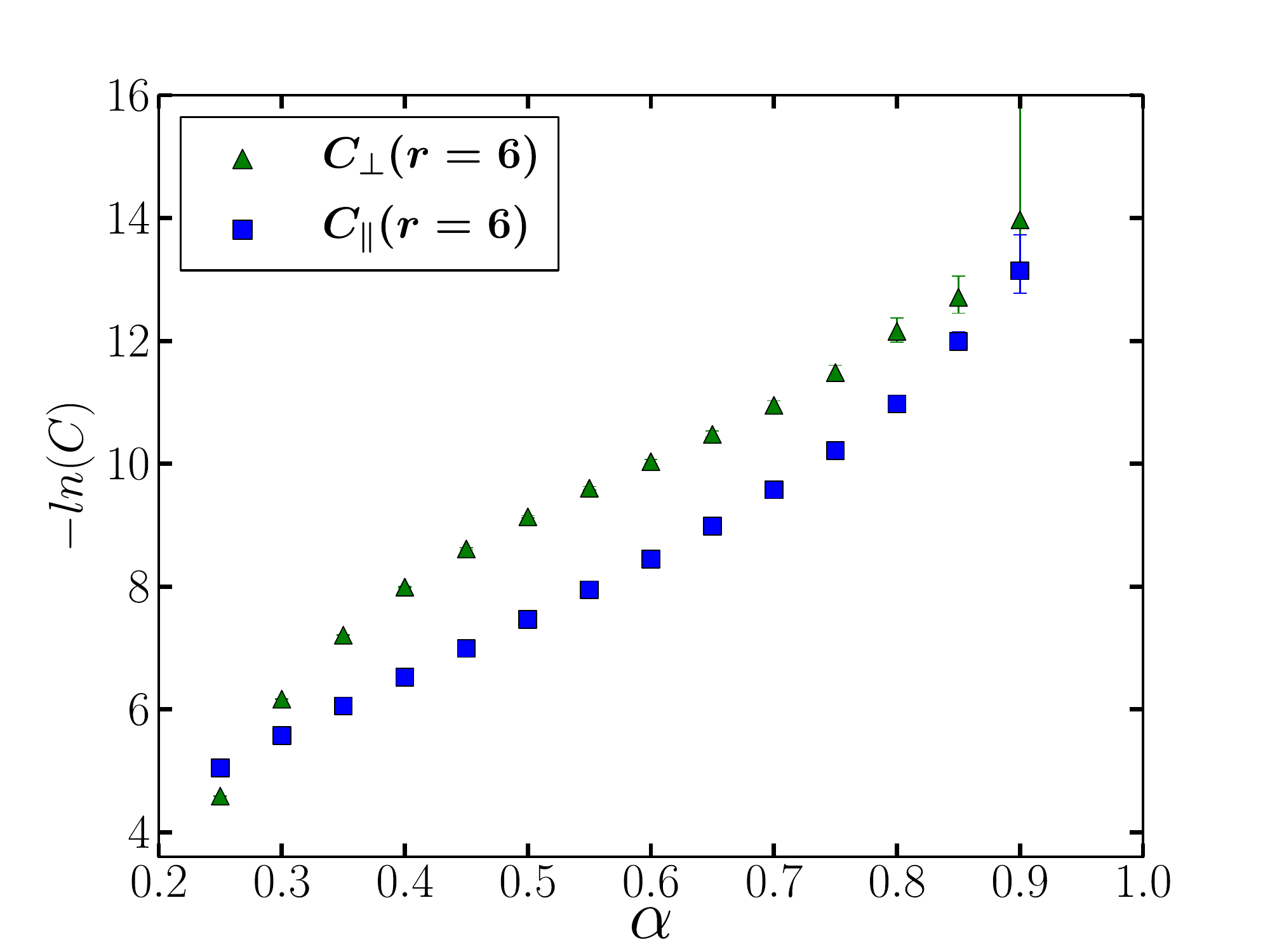}
\caption{(Color online) Negative natural logarithm of the dimer-dimer correlation at six lattice spacings apart on an $18\times18$ lattice for the ground state of Eq.~(\ref{eq:inter}) as a function of the parameter $\alpha$. $\parallel$, $\perp$ refer to dimers parallel to each other or offset by $60^{\circ}$, respectively.}
\label{fig:cor}
\end{figure} 

The TEE begins to diverge from the theoretical value, when the bipartite region length scale $w$ is on the order of 10 times the correlation length $\xi$ [for the current R\'{e}nyi index $n=2$). For the interacting wave function (ground state of Eq.~(\ref{eq:inter})], the dimer-dimer correlation length does not diverge, since the transition is first order, however, it grows compared to the RK values. Figure~\ref{fig:cor} shows the negative natural logarithm of dimer-dimer correlation at 6 lattice spaces apart---the length scale of the boundary bipartitions $w$ on an $18\times18$ lattice (the lattice size for the results on Fig.~\ref{fig:NFPLQ}). We use the negative natural logarithm of the correlation function at a characteristic length $w (= 6)$, to estimate $w/\xi$ , where $\xi$  is an effective correlation length. The divergence from $-\ln 2$, beginning around $\alpha=0.5$, corresponds to a value of $8$ in Fig.~\ref{fig:cor}.  
Expressed in this way, the value can be compared to the results of St\'{e}phan \emph{et al}.,~\cite{Stephan2011} where a deviation can also be seen at similar values [since $t \sim 1/\xi$,~\cite{Fendly} and deviations begin near $L t \sim O(10)$]. 
Taken together, these results suggest that, at least for dimer systems, the TEE is quite sensitive to finite correlations, requiring bipartition length scales up to $O(10)$ times the correlation length. If such scaling holds more generally, it will constitute an important limitation on the use of numerics to extract the TEE.

\section{Corner contributions to entanglement entropy}
\label{sec:corners}
\begin{figure} 
\includegraphics[width=.47\textwidth]{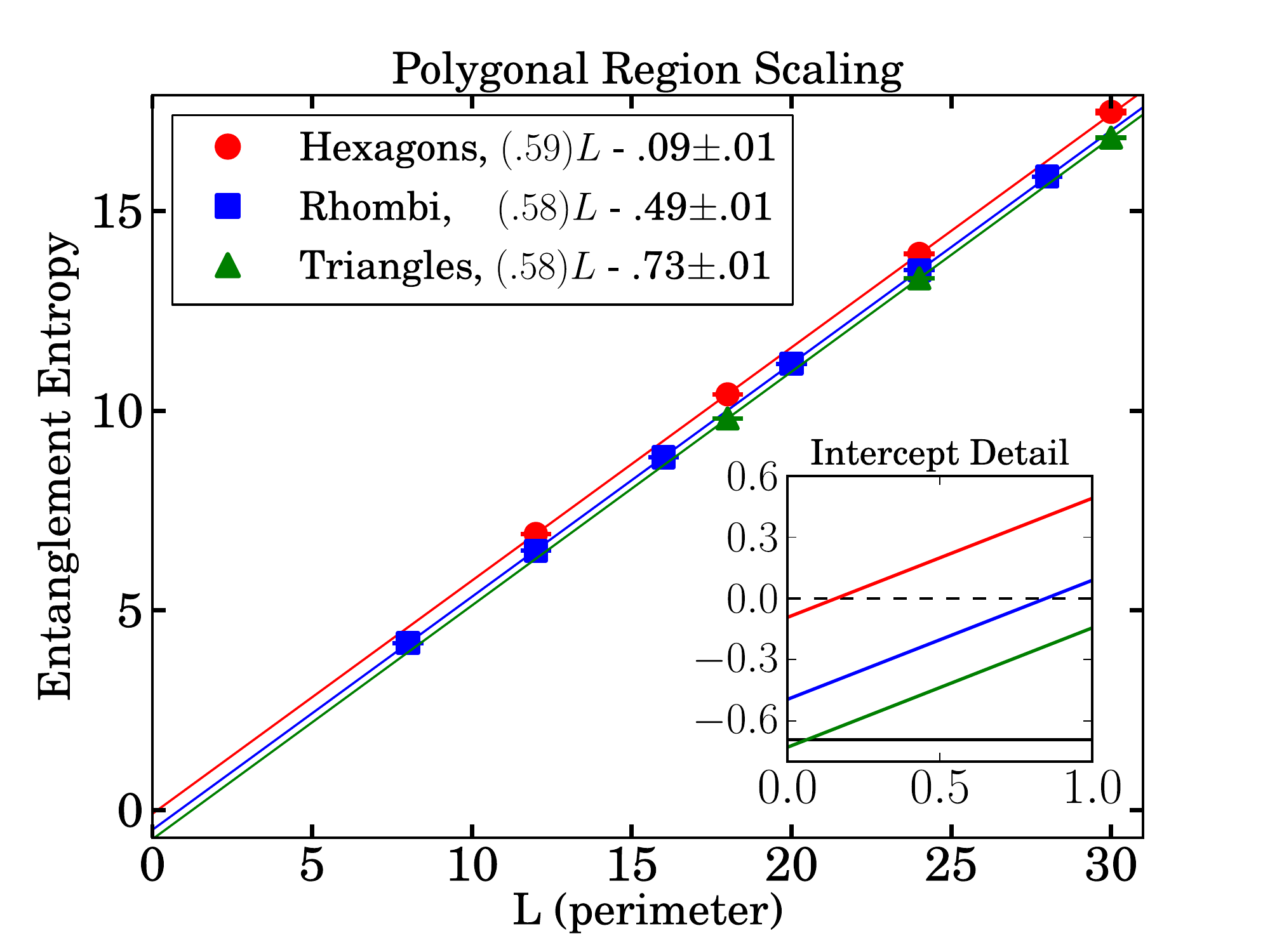} 
\caption{(Color online) Entanglement entropy and best fit line as a function of polygonal region perimeters, $L$, showing offsets for different polygonal regions. Fits include total constant term; for comparison the expected $\gamma =\ln 2 = 0.693$ is shown (solid line in the inset).}
\label{fig:scalingfits}
\end{figure}

As noted earlier, bipartitions with corners contribute a non-universal constant, $\kappa$, to the entanglement entropy. The presence of corner shifts can be readily seen in the scaling of various polygonal regions, in particular, of the hexagons, triangles, and rhombi shown in Fig.~\ref{fig:scalingfits}. %
The offset appears to be a constant shift. This is more precisely measured by computing differences of bipartitions as discussed below. 

Corner effects have been studied for the integer quantum Hall wave function.~\cite{Rodriguez} In other topological phases there have not been any studies on corner contributions that we are aware of, although in Ref.~\onlinecite{Stephan} also the potential for non-universal constant corner contributions away from the critical point was noted.  Our results are consistent with a total contribution $\kappa$ of the form $\kappa = \sum\limits_{i} a_i n_i$ as found for quantum Hall  systems in Ref.~\onlinecite{Rodriguez}, where $i$ labels the types of corner, $n_{i}$ the number of corners of type $i$ while $\{a_i\}$ are coefficients to be determined. With this form, $\kappa$ exactly cancels for the Kitaev-Preskill construction if, as required by the symmetry of entanglement entropy $a_i=a_{\tilde{i}}$, where $\tilde{i}$ is the complement of $i$ (the angle which combines to form a closed circle). Apart from the numerical values, the most important result of this section is that corner effects do saturate to constant shifts, allowing for the extraction of the TEE by the approach of canceling regions. 

\begin{figure} 
\includegraphics[width=.125\textwidth]{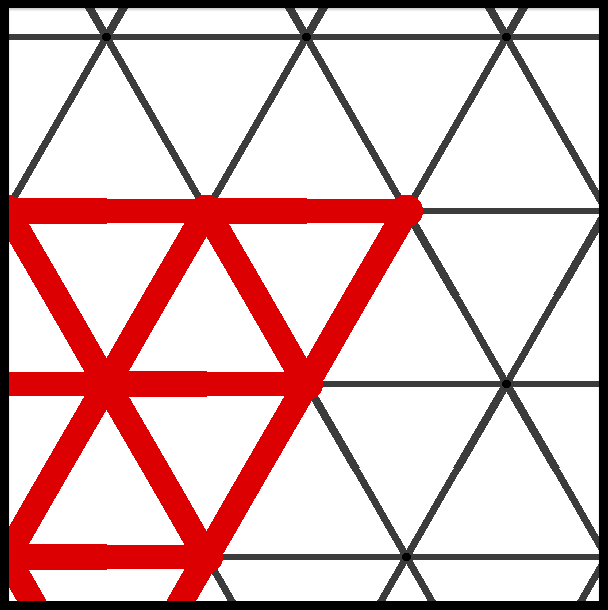} 
\includegraphics[width=.125\textwidth]{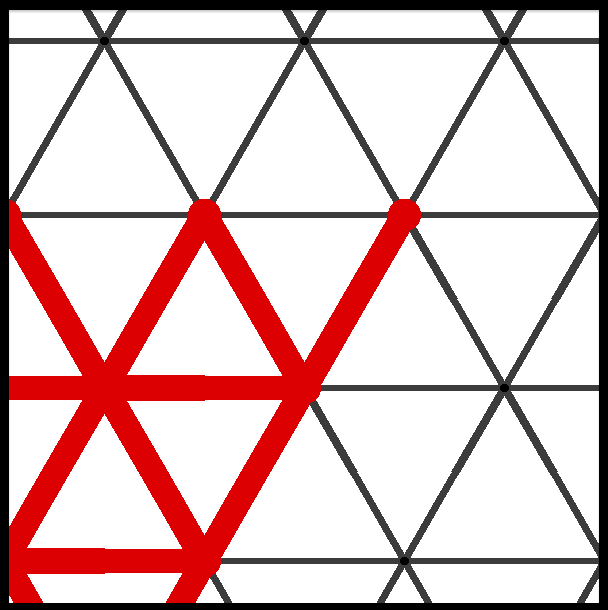} 
\includegraphics[width=.125\textwidth]{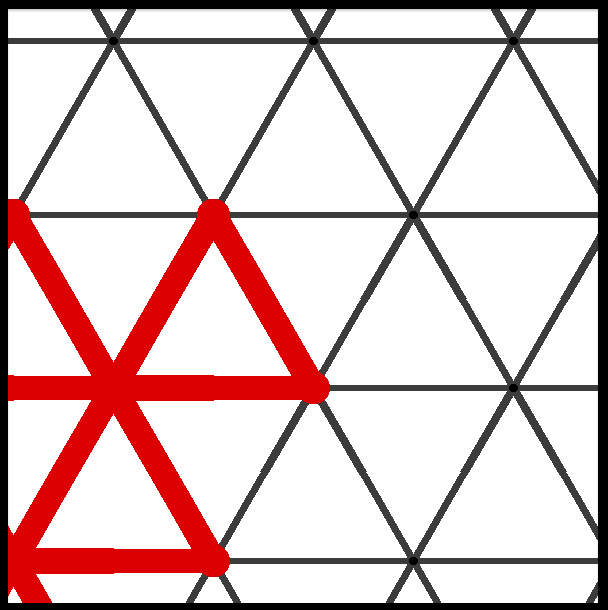}
\vspace{4pt}
\\
\includegraphics[width=.125\textwidth]{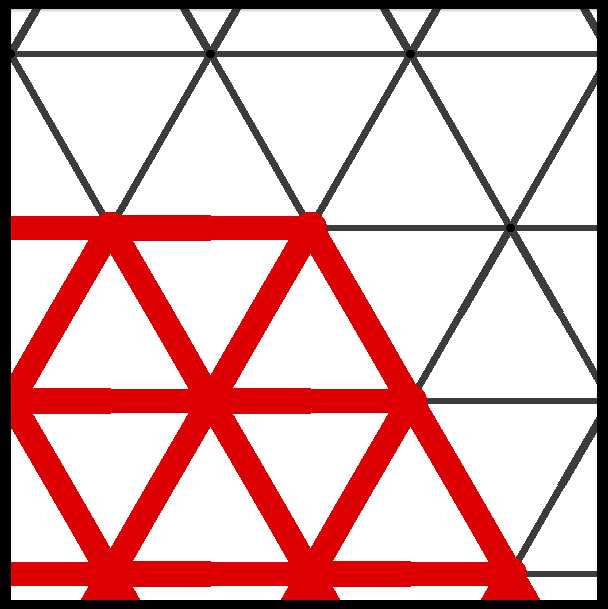} 
\includegraphics[width=.125\textwidth]{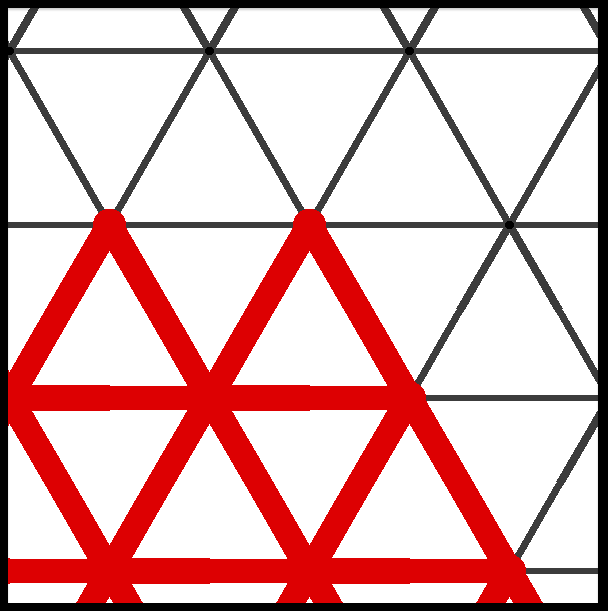} 
\includegraphics[width=.125\textwidth]{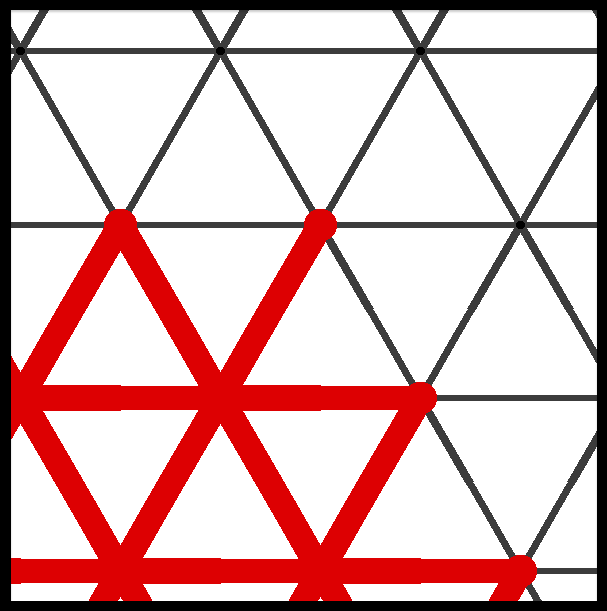} 
\quad
\caption{(Color online) Six lattice corners. Type-$\I$ corner (top), left to right, $\alpha$, $\beta$, $\sigma$. Corresponding type $\II$ on bottom.}
\label{fig:corners}
\end{figure}

Since the linear shifts in Fig.~\ref{fig:scalingfits} include the topological term, the triangles actually have the smallest shift, despite having sharper corners. More precise results confirm this. To see why sharper corners can have a small effect, first note that unlike the continuum, the types of corners $i$ are not solely determined by angle. Microscopically, on the triangular lattice there are a total of six types of corners (plus 
complements) labeled as $i \in$ $\I\alpha$,  $\I\beta$, $\I\sigma$, $\II\alpha$,  $\II\beta$, and $\II\sigma$. The roman numeral refers to the angle ($\I= 60^\circ$, $\II=  120^\circ$) and the greek letter as to whether incoming links are included or 
not, as shown in Fig.~\ref{fig:corners}.
\begin{table} 
\caption{Corner shifts $a_i$. The estimate is computed from the analysis described in the Appendix and Supplemental Material. See text and Fig.~\ref{fig:cornfits} for details of the Monte Carlo runs employed.}
\label{tab:corner}
\newcolumntype{.}{D{.}{.}{12}}
\newcolumntype{u}{D{.}{.}{6}}
\newcolumntype{t}{D{.}{.}{4}}
\newcolumntype{s}{D{.}{.}{0}}
\begin{tabular}{|cutst|}
\hline
Corner  &  \multicolumn{1}{|c|}{Estimate} & \multicolumn{3}{|c|}{Measured}  \\
\hline

 $\I{\alpha}$  &   -0.02& 0.008 &\pm&0.002 \\ %
 $\II{\alpha}$  &  0.09  & 0.0946&\pm &0.0001  \\
$\I{\beta}$   &  -0.26 &  -0.2592&\pm &0.0004 \\ 

$\II{\beta}$  &  -0.02   & -0.0011 &\pm& 0.0004\\ 
$\I{\sigma}$   &  -0.62 & -0.677&\pm &0.002 \\ 
$\II{\sigma}$  &  -0.23   &  -0.238 &\pm &0.002\\ %
\hline
\end{tabular}
\end{table} 
\begin{figure*} 
\includegraphics[width=.32\textwidth]{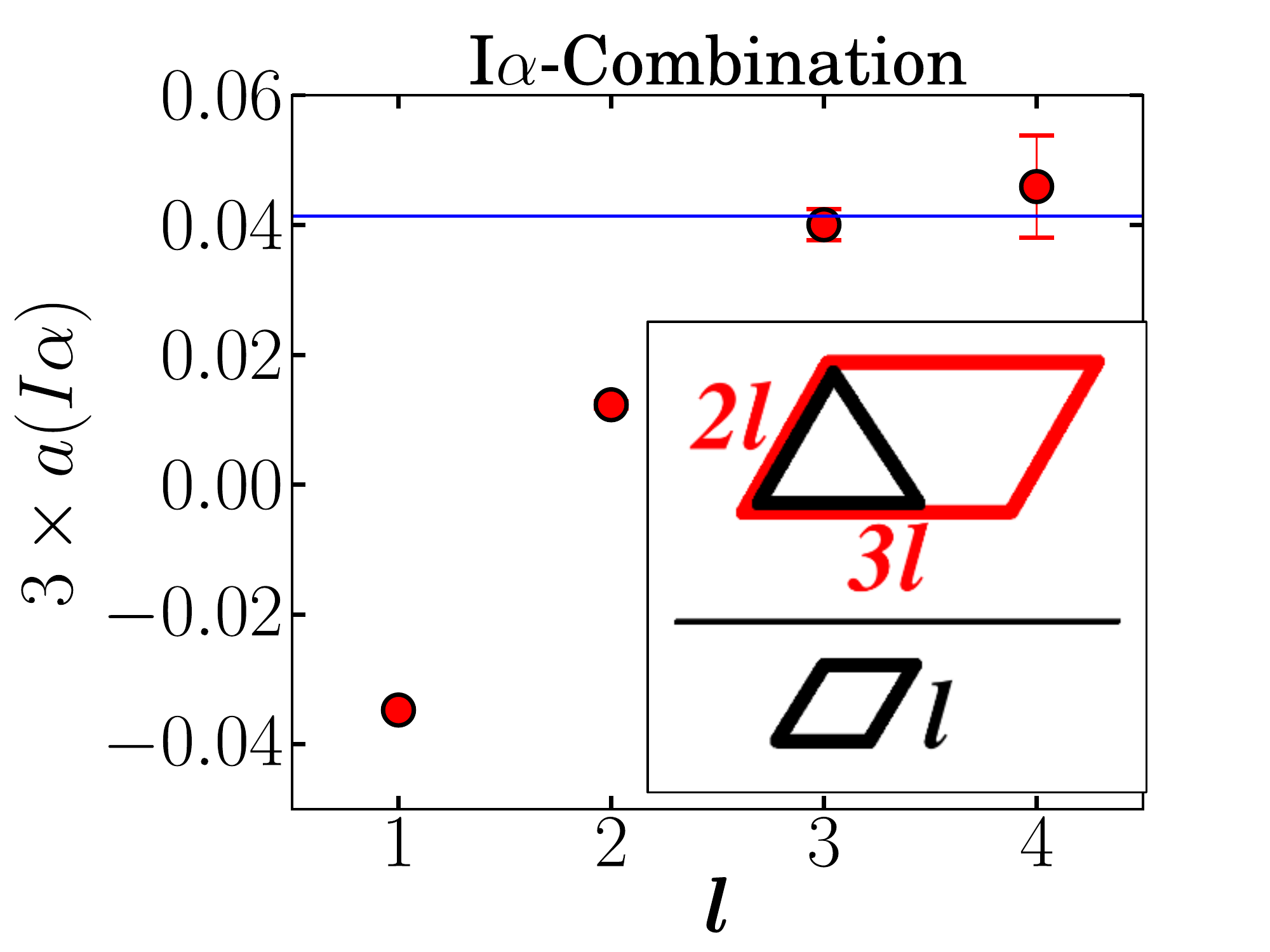} 
\includegraphics[width=.32\textwidth]{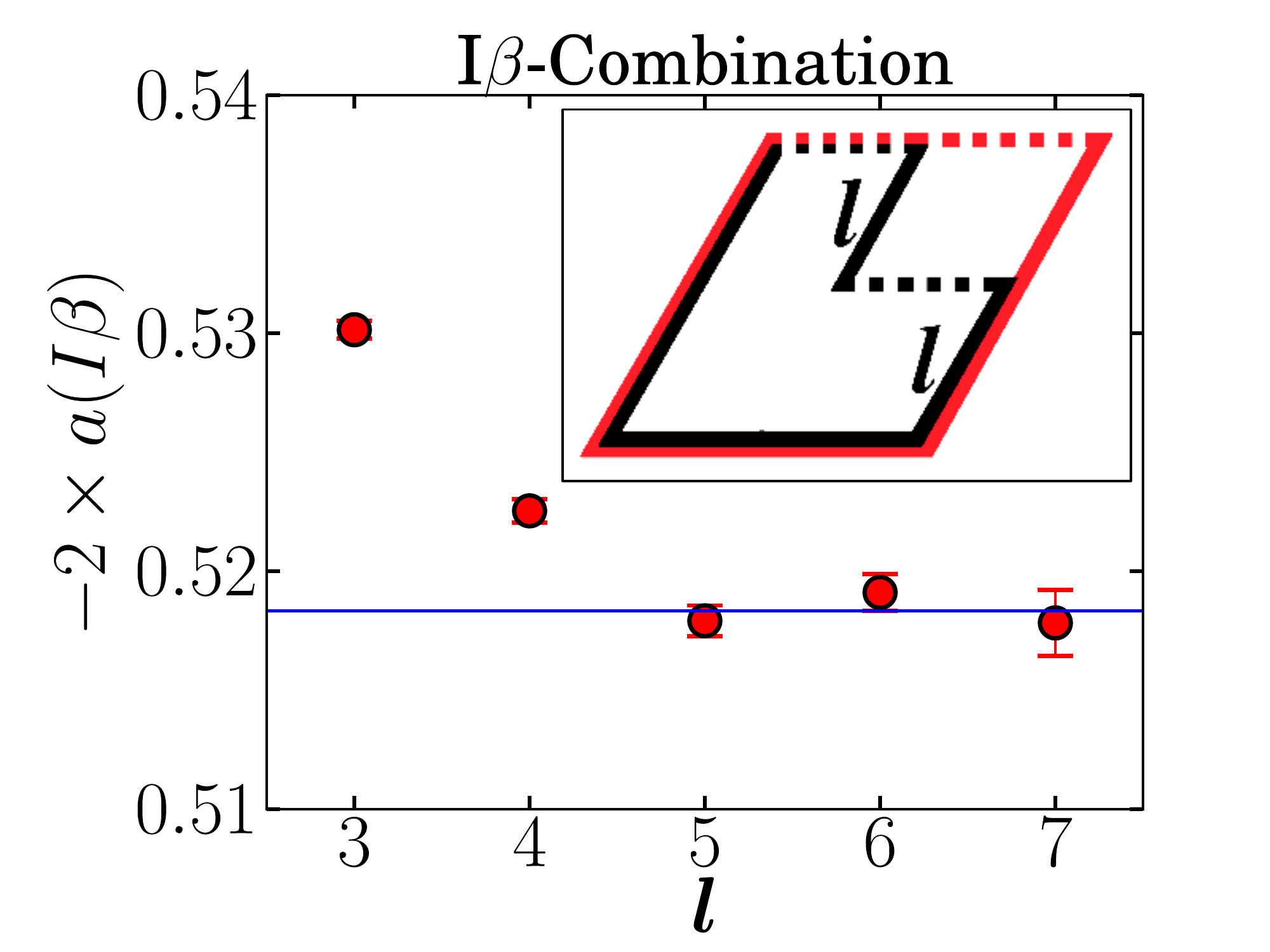} 
\includegraphics[width=.32\textwidth]{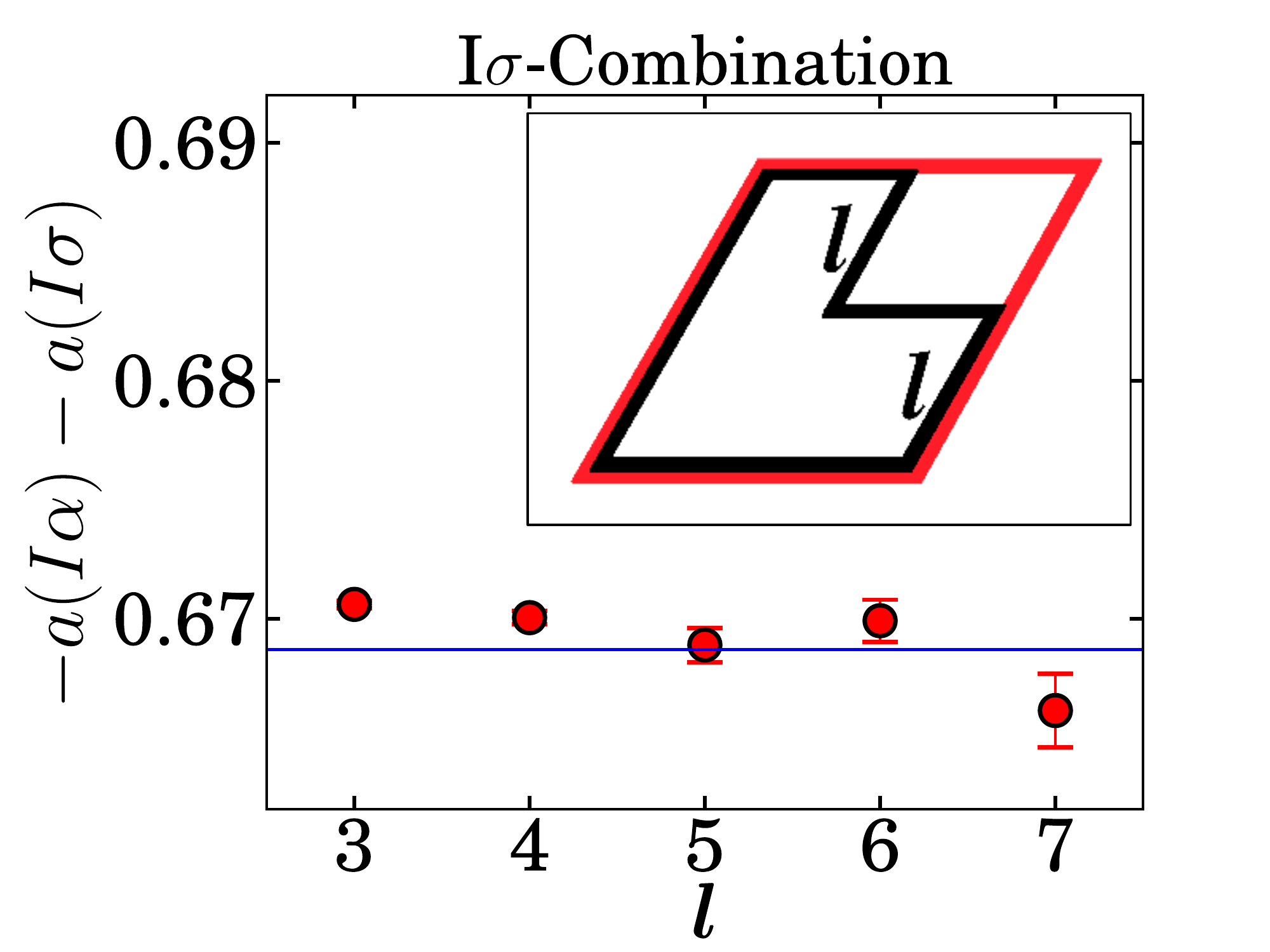} \\
\includegraphics[width=.32\textwidth]{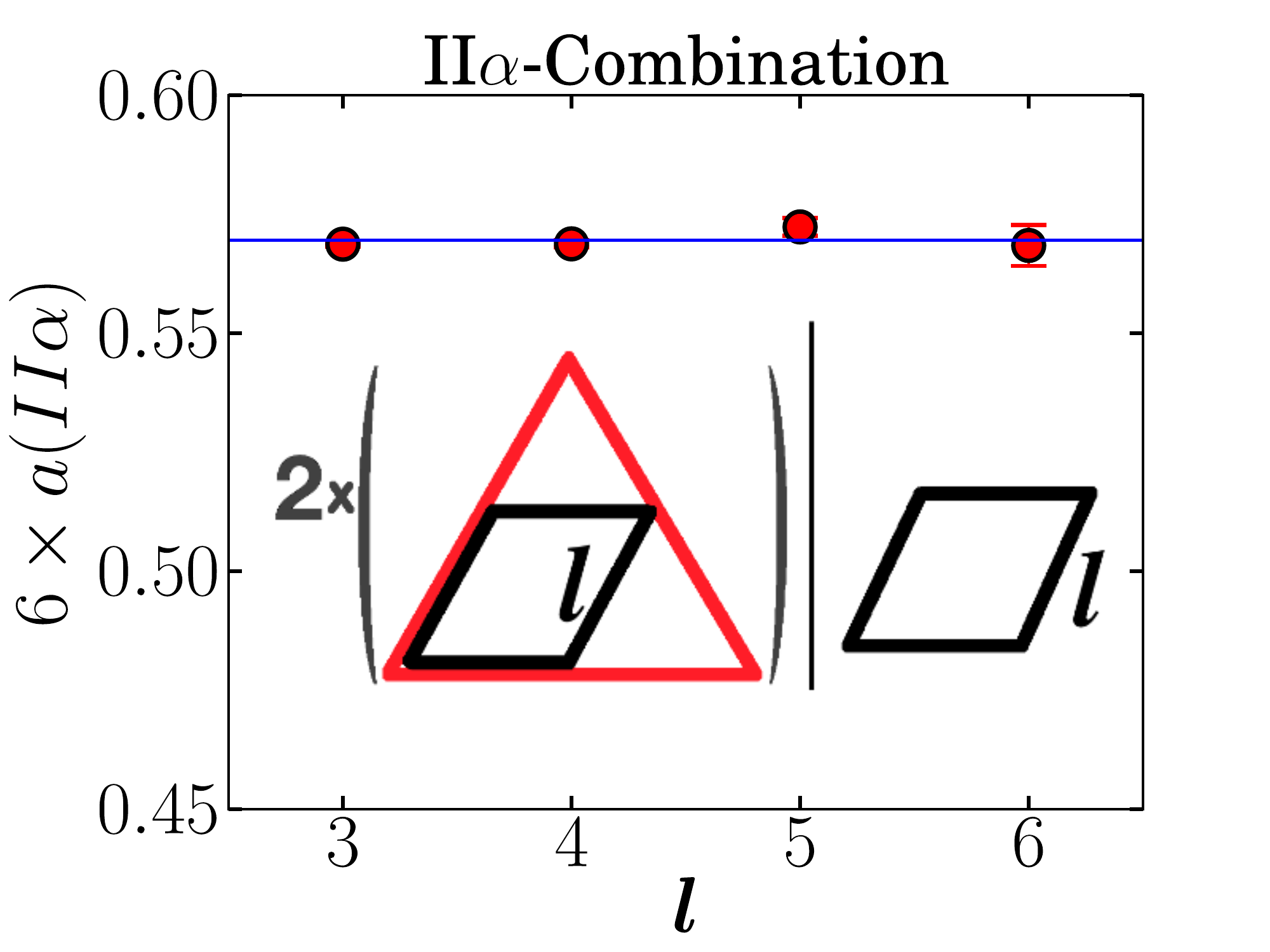}
\includegraphics[width=.32\textwidth]{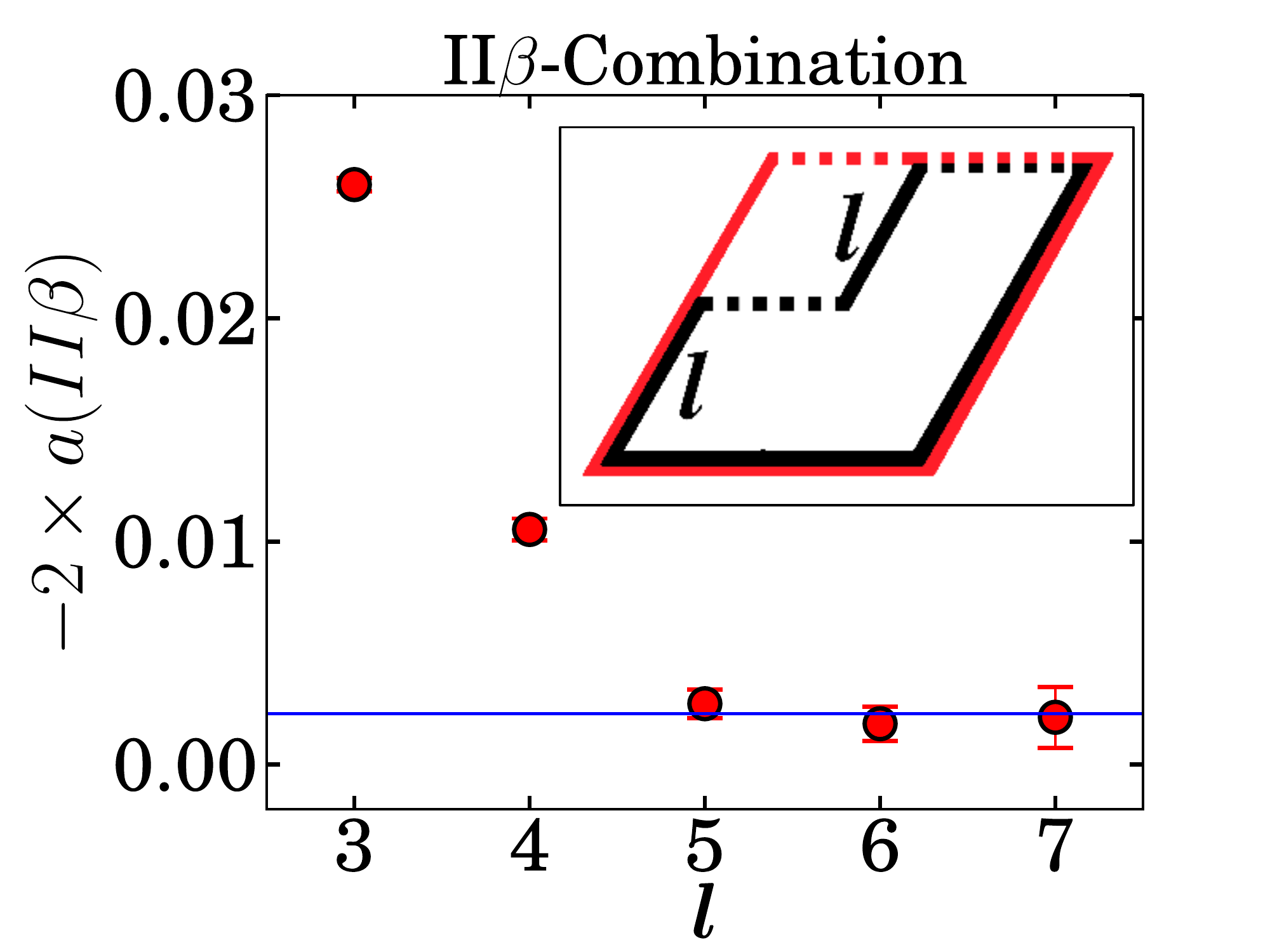}
\includegraphics[width=.32\textwidth]{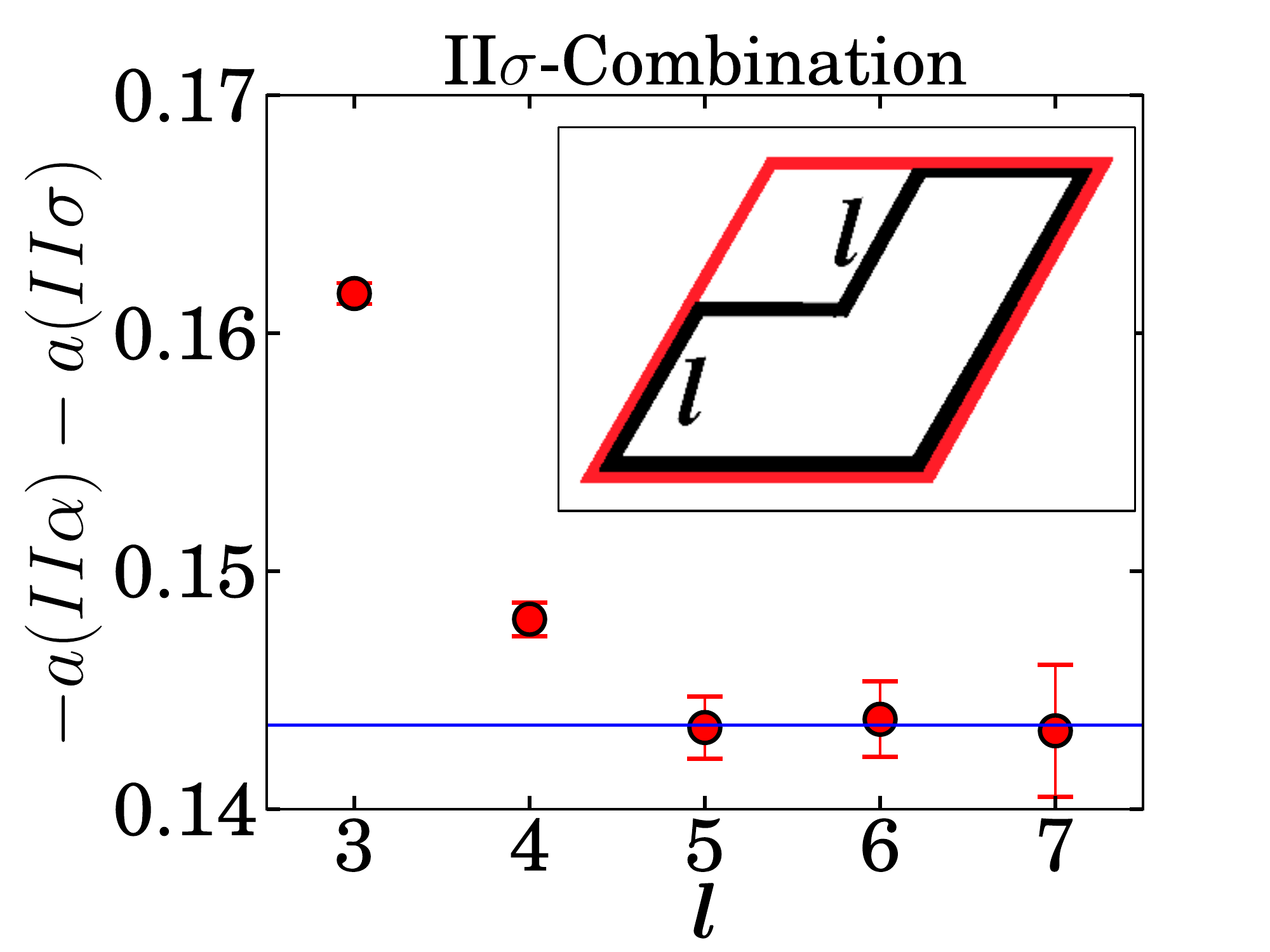}
\caption{(Color online) Bipartition geometries used to extract corner contributions. Entropy differences are computed between the black and red regions. The dotted line represent sides with ``missing links'' used to form $\beta$ corners. The $\I\alpha$ and $\II\alpha$ combinations involve pairs plus an extra region, thereby requiring a subtraction of the TEE (taken to be $\ln 2$). The blue horizontal lines are the values used for table~\ref{tab:corner}.}
\label{fig:cornfits}
\end{figure*} 
Measured values of the shifts $a_i$ are presented in Table~\ref{tab:corner}. Estimates are obtained from a mean-field-like analysis presented in the Appendix and Supplemental Material.~\cite{SM} Two important points can be deduced from the analysis that give insight into the form of $\kappa$ and also the original question, i.e., why sharp corners have a small contribution. First, the corner terms can be thought of most naturally as part of a generalized linear scaling with boundary vertices $\{v\}$, $S_{A} \sim \sum_j \alpha_{j} \ell_j = \sum_j \alpha_{j} \ell$ for $j\in \{v\}$, where $\ell$ is the lattice spacing. When all $\alpha_j$ are equivalent 
($\alpha_j =\alpha$),
 as in a straight side, this becomes a linear scaling $= L_{A} \alpha$. Then the corner shift can be understood as single dislocation 
 $\alpha \rightarrow \alpha_{i}$, so that $a_i = \Delta\alpha= \alpha_{i}-\alpha$. Second, to first order, the scaling $\alpha_{j}$ is determined by \emph{the number} of links in region $A$ versus $B$ (the complement) that meet at boundary vertices. For side vertices away from corners these are 4 and 2, therefore corners $\I\alpha$ and $\II\beta$ which have the same incoming links have the smallest shift, followed by $(3, 3)$ which adds entropy for $\II\alpha$, then $(1, 5)$ for $\I\beta$ and $\II\sigma$, and $(0, 6)$ for $\I\sigma$.
These values are all in agreement with the ordering shown in Table~\ref{tab:corner}.

Several Monte Carlo runs were used to extract the measured $a_i$.  The types $\I\alpha$ and $\II\alpha$ were taken by combining the information from: (1) the linear scalings in Fig.~\ref{fig:scalingfits} after removing the TEE entropy (assuming it is $\ln 2$), (2) the differences between those polygons (note \emph{any} combination of those gives only $2a(\II{\alpha})-a(\I{\alpha})$), and (3) specially constructed combinations called ``$\I\alpha$ and $\II\alpha$ combinations" shown in Fig.~\ref{fig:cornfits}.  The measured $\I\beta$ and $\II{\beta}$ are easily obtained solely from the L-shaped combination shown in Fig.~\ref{fig:cornfits} (``$\I\beta$ and $\II\beta$ combinations"); similarly for $\I\sigma$ and $\II{\sigma}$.

In Fig.~\ref{fig:cornfits}, the most important feature is that corner effects do indeed saturate to constant shifts, rendering constructions like the Kitaev-Preskill approach useful for extracting the TEE.

\section{Conclusions}
\label{sec:conclusions}

We have demonstrated that Monte Carlo techniques provide a viable method for the numerical calculation of the TEE of the ground states of generalized RK points. In addition to confirming existing results, we have also been able to apply the method to a generalized RK wavefunction and to thereby investigate the behavior of TEE approaching a quantum phase transition. Our results suggest that the TEE is indeed a robust indicator of topological order in the thermodynamic limit throughout the quantum liquid regime. However, we also find a strong dependence on correlations that requires bipartitions with side lengths of order at least ten times the correlation length. If this estimate applies generally, it implies that numerics will be severely constrained in the vicinity of a second order phase transition. These results are therefore an important guide for future quantum Monte Carlo studies.

The third aspect that we have been able to study here is the nature of corner contributions.
First the magnitude of corner shifts is of order $\gamma$ so their effects must be controlled either by canceling them out or by dealing exclusively with smooth boundaries. On a lattice with periodic boundary conditions, however, these are not topologically trivial. For example, difficulties in the linear extrapolations of Ref.~\onlinecite{Furukawa} may have been due to the effective variation of edges associated with the radial-like region definitions.
Second, at least for the RK point, we can verify that corner shifts do indeed saturate to constant values, thereby making possible constructions which cancel corners linearly. It remains to be seen whether this is also the case for more general wavefunctions. 

Our results are important for future work employing quantum Monte Carlo calculations and for conclusions about the practical utility for TEE as a probe of topological order. In this regard, 
the present results allow us to conclude that the TEE is a useful tool as long as correlation lengths are small enough in at least one point of the topologically ordered liquid phase. While the only other known example of 
the behavior of the TEE on approaching a quantum phase transition does suggest a similar behavior,~\cite{Stephan} further numerical studies of TEE in exotic phases of matter should be of great value.

\emph{Note added.} We recently became aware of similar work by Pei \emph{et al}.~\cite{Pei}

\acknowledgements
This work was supported by NSF Grant No. PH4-0803429.

\appendix


\section{Estimate of corner effects.}
Using Kasteleyn matrices,~\cite{Kasteleyn, Fisher} an estimate for corner effects can be obtained. Kasteleyn showed that the number of dimer coverings on a lattice is given by the Pfaffian of a matrix $K_{ij}$, with $i$, $j$ labeling lattice \emph{sites} and $K_{ij}\pm1$ (or suitable weights in generalizations) for nearest neighbors and zero otherwise; the sign is determined according to a convention of  arrows placed on the lattice. We refer readers to Refs.~\onlinecite{Fendly, Stephan}. Only two facts will be used for the following arguments. First, $Z$ is equivalent to the number of coverings and equals the Pfaffian of $K$ (written $\text{Pf}(K)$). Second, the square of the Pfaffian is the determinant for an even-dimensional matrix: $[\text{Pf}(K)]^2 = \text{Det}(K)$.

\begin{figure} 
\includegraphics[width=.3\textwidth]{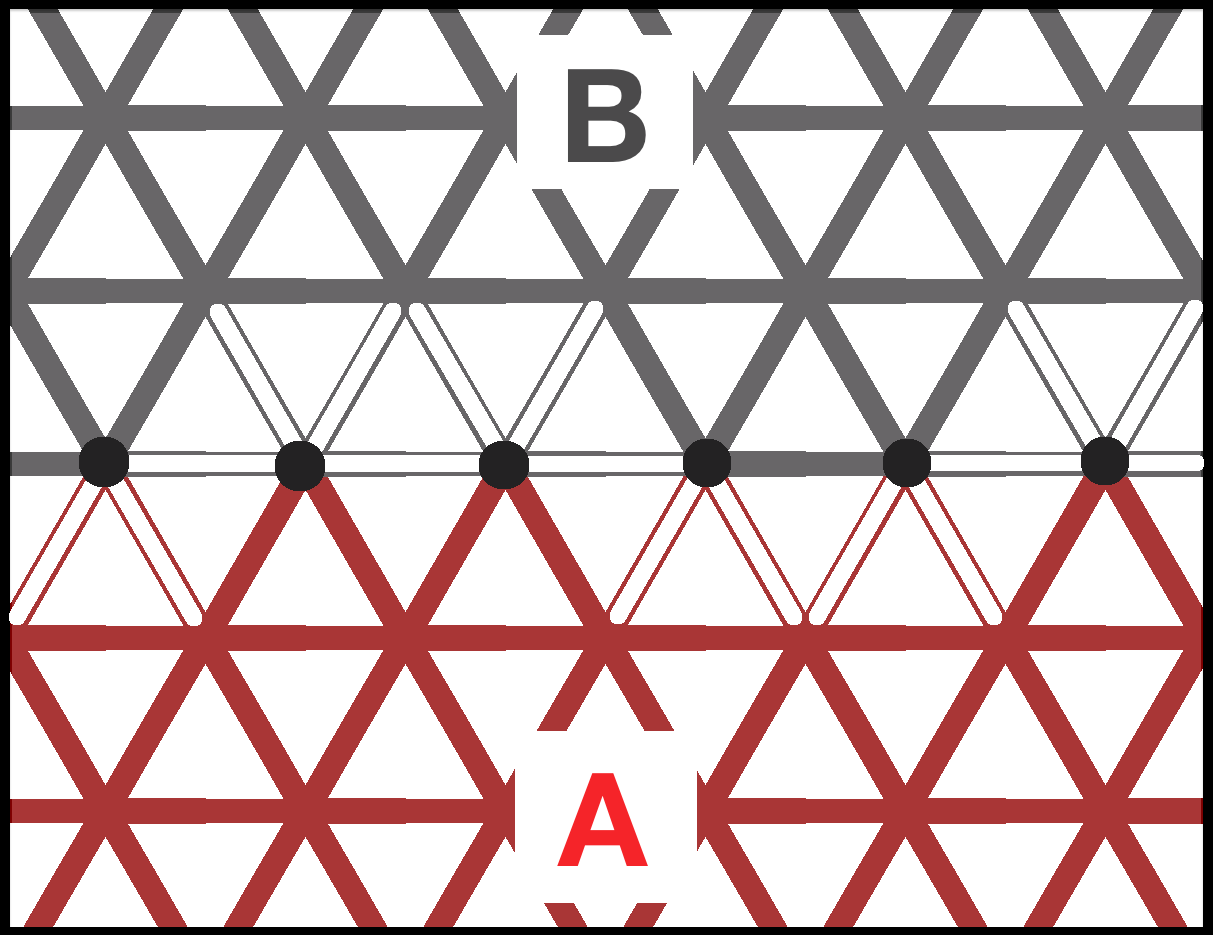} 
\caption{(Color online) Boundary vertices in black shown taking values $\{ v\} = \{ baabba\}$ (periodic boundary conditions are not assumed). The lattice to be included in $K$-$E$ is shown by the darkened links.}
\label{fig:kast}
\end{figure}

For a configuration to be swappable, one simply needs to satisfy the dimer constraint along boundary vertices is equivalent to $\{v\}$, which are vertices that have links  
belonging to both $B$ and $A$. For each boundary vertex in each configuration, a dimer can be on a $B$ link or an $A$ link, which we will consider as a \emph{``side parity''}  labeled as either $v=a$ or $ b$ if the vertex has a dimer in $A$ or $B$. Then swappability simply imposes that for every boundary 
vertex, the dimer in copy 1 is in the same \emph{side} as the dimer in copy 2 ($v_1=v_2$). Using a similar notation to that of Ref.~\onlinecite{Stephan}, the number of swappable configurations can be written as:
\begin{equation}
Z_{\SWAP}=\sum\limits_{\{v\}} Z_{1}|_{\{v\}}\cdot  Z_{2}|_{\{v\}}, 
\end{equation}
where $Z_{1, 2}|_v$ is the set of dimer coverings in copy 1(2) given a set of $A$, $B$ occupations, or side parities of $\{ v\}$; the sum runs over the combinations specifying $a$ or $b$ for the number of boundary vertices, is equivalent to $N_v$. Note that  $Z_{1}|_{\{v\}} = Z_{2}|_{\{v\}}$, and each can be written as a Pfaffian, with certain constrained links removed. The links to be removed are those which touch a vertex $v$ on the side $A$ if $v=b$, or on the side $B$ if $v=a$ (the Pfaffian will then generate combinations on the occupied side).
An example is shown in Fig.~\ref{fig:kast}. An ``exclusion matrix'' $E$ can be defined so that for a \emph{removed} link from site $r$ to $s$, $E_{rs}=K_{rs}$ and it is zero otherwise, so that:
\begin{equation}
\begin{split}
Z_{\SWAP}= \sum\limits_{\{v\}} [\text{Pf}(K-E|_{\{v\}})]^{2} = \sum\limits_{\{v\}}\text{Det}(K-E|_{\{v\}})
\end{split}
\end{equation}
with $K$ and $E$ now referring to only a single copy. 
Finally from Eq.~\ref{eq:swap} the entanglement entropy can be written as:
 \begin{equation}
\begin{split}
S_A= -\ln \frac{Z_{\SWAP}}{Z_1^2}  = -\ln \sum\limits_{\{v\}}\frac{\text{Det}(K-E|_{\{v\}})}{\text{Det}(K)} 
\end{split}
\end{equation}
which matches the $n=2$ R\'{e}nyi entropy in Ref.~\onlinecite{Stephan}. Next we can employ a perturbation theory of matrices\cite{Fisher1963} and the trace identity,
\begin{equation}
\begin{split}
\frac{\text{Det}(K-E|_{\{v\}})}{\text{Det}(K)}&= \frac{\text{Det}(K)\text{Det}(1-K^{-1}E|_{\{v\}})}{\text{Det}(K)} 
\\
= \exp[\text{Tr}\ln(&1-K^{-1}E|_{\{v\}})] \\
= \exp[\text{Tr}(-&K^{-1}E - \frac{1}{2}K^{-1}EK^{-1}E\dots)],
\end{split}
\end{equation}
with $E|_{\{v\}}$ implied in the last line.  $K^{-1}$ can be diagonalized in Fourier space.\cite{Fendly} It decays exponentially in vertex separation; for nearest neighbors $i, j$ $K^{-1}_{i, j}=\pm1/6$, it vanishes for next to nearest, and is $\sim0.02$ for a three-link separation. Because the number of terms contributing to the trace also grows, the series does not converge as fast as $K^{-1}\sim 1/6$, but roughly $\sim 1/2$, so that at least these two terms should be kept, and we expect the estimates to be accurate to the order of $10\%$ if we ignore higher order terms of the expansion. Ignoring those terms and making the definitions $ T_1|_{\{v\}} \equiv \text{Tr}(-K^{-1}E|_{\{v\}} ) $ and $T_2|_{\{v\}} \equiv  \text{Tr}(- \frac{1}{2}K^{-1}E|_{\{v\}}K^{-1}E|_{\{v\}})$, the resulting form is
 \begin{equation}
S_A = -\ln \sum\limits_{\{v\}}\exp{(T_1|_{\{v\}}+T_2|_{\{v\}})}.
\label{eq:S_A}
\end{equation}
The sum ranges over the $a, b$ choices for each vertex.

If $T_{1(2)}$ depended only on the number of $a$ versus $b$ vertices Eq.~\ref{eq:S_A} could be written as a binomial expansion or Gaussian in the large $N_v$ limit. However, in general $T_{1(2)}$ depend on the details (ordering) of a particular set of $a, b$ choices of $\{v\}$. Specifically, the first term $T_1= (K^{-1})_{ij}(E|_{\{v\}})_{ji}$ is only non-zero for $i,j$ nearest neighbors of the ``excluded lattice''. It is evaluated as $-\frac{1}{6}\sum_{i} z_i \equiv T_1$, where the sum is over all vertices $i$, and $z_{i}$ is the number of nearest neighbors for the $i$th vertex (of the excluded lattice). $T_1$ can be readily summed to $-\frac{2}{6}{\times\text{(number of removed links})}$.  This value depends on the number of $a$ versus $b$ values and also the ordering. For example, the number of links in the``excluded lattice'' increases by 2 for every $v=b$, but may increase by 3 or 4 for $v=a$, depending on whether it follows an $a$ or $b$, respectively. Therefore to proceed, one can define a mean-field contribution as $-\frac{2}{6} \times2$ per $v=b$ vertex while $-\frac{2}{6} \times[ 3P + 4(1-P)]$ per $v=a$ vertex where $P$ is the likelihood of finding a $v=a$ neighbor, which can be determined self-consistently.
The second term $T_2$ can be put in a similar form provided one ignores $K^{-1}$ terms connecting three links or more, as these contain a factor of 0.02. Then $T_2$ can be shown to be $-\frac{1}{72}\sum_{i} [z_i^2+ z_i(z_i-1)]$. One can also make a mean field estimate for the average contribution to $T_2$ by considering combinations of neighboring preceding vertices.
Note, that the form of these terms as a sum suggests the potential for the generalized linear scaling mentioned in Sec.~\ref{sec:corners}, $S_A \sim \sum_i \alpha_i \ell_i$. 

If we call the average contribution to $T$ per $v=a$ vertex on a straight side, $\equiv \tau_{side,a}$ and similarly per $v=b$ vertex  $\equiv \tau_{side,b}$, then given a combination of vertices $\{v\}$ (along a side) having $n_a$, $v=a$ vertices, $T\equiv T_1+T_2$  can be written in terms of these average values: 
\begin{equation}
T = n_a\tau_{side,a}+(N_v-n_a)\tau_{side,b}.
\end{equation}
Put into this form (depending only on the \emph{number} of $a$'s, $n_a$, in $\{v\}$), the sum [Eq.~\ref{eq:S_A}] can be written as a binomial expansion and approximated as the Gaussian integral in the large-$N_v$ limit:
\begin{widetext} 
\begin{equation}
\begin{split}
S_A \approx& -\ln \int d{n_a} \sqrt{\frac{2}{\pi N_v}}\exp\left(-\frac{2 (n_a-N_v/2)^2}{N_v} +N_v\ln(2) +\tau_{side,b}(N_v-n_a) + \tau_{side,a} n_a\right) \\
= &N_v\left(-\ln(2) -\frac{\tau_{side,b}+\tau_{side,a}}{2}-\frac{(\tau_{side,a}-\tau_{side,b})^2}{8}\right) \equiv \sigma_A(N_v),
\end{split}
\label{eq:gauss}
\end{equation}
\end{widetext}

Finally a corner shift for a corner of type $c$, can be extracted from the difference $\Delta S_A(c) \equiv$  $S_A$($N_v-1$ side vertices $+ c$ ) $-S_A$($N_v$ side vertices); with the exception of $\I\sigma$ discussed below. One can think of the first term as a binomial(Gaussian) sum over $N_v -1$ non-corner terms multiplied by a factor of $\times(e^{\tau_{c,a}}+e^{\tau_{c,b}})$ which generates the last two possibilities at the corner with similar mean-field contributions for corner vertices: $\tau_{\I\alpha,a}, \tau_{\I\alpha,b}, \tau_{\I\beta,a}, \tau_{\I\beta,b}$, etc.~\cite{SM} Then we can write:
\begin{equation}
\Delta S_A = \sigma(N_v-1)-\ln(e^{\tau_{c,a}}+e^{\tau_{c,b}})-\sigma(N_v),
 \end{equation}
 or,
\begin{equation}
\begin{split}
\Delta S_A = &\left(\ln(2) +\frac{\tau_{side,b}+\tau_{side,a}}{2}+\frac{(\tau_{side,a}-\tau_{side,b})^2}{8}\right) 
\\&-\ln(e^{\tau_{c,a}}+e^{\tau_{c,b}}).
\label{eq:shift}
\end{split}
\end{equation}
For the case of $\I \sigma$, at the ``corner'' there is no boundary vertex. To compute the shift in this case, instead two vertices are removed, and the factor $\tau^*_{\I\sigma,a}$ refers to a substitution of the vertex just before the corner. So the above formulas can be used with $\tau^*$'s and the first term (enclosed in parentheses) multiplied by 2.

Insertion of estimates for the $\tau$'s (see the Supplemental Material~\cite{SM}) yields the results given in 
Table~\ref{tab:corner} which are in good agreement with the measured values. Also, note that the 
term multiplying $N_v$ in $\sigma_A(N_v)$ (Eq.~\ref{eq:gauss}) is an estimation of the slope of the entanglement entropy. The value for the slope obtained by plugging in the mean-field values is $0.51$, which is on the order of $10 \%$ different from the fitted slope values near $0.58$. The relative success of the substitutions suggest that the corners can be seen 
explicitly in the generalized linear scaling $\sum_i \alpha_i \ell_i$ noted above and in Sec.~\ref{sec:corners}.

\bibliographystyle{apsrev4-1}
\bibliography{CS}

\clearpage


\section*{Supplementary material}

# Supplemental Material for Entanglement Entropy at Generalized RK Points of Quantum Dimer Models

Alexander Selem[1,2],[*] C. M. Herdman[1,2,3],[†] and K. Birgitta Whaley[1,3]
[1]Berkeley Quantum Information & Computation Center,
University of California, Berkeley, CA 94720, USA
[2]Department of Physics, University of California, Berkeley, CA 94720, USA and
[3]Department of Chemistry, University of California, Berkeley, CA 94720, USA

## I. MEAN-FIELD APPROXIMATIONS FOR CORNER EFFECTS

We can estimate on average the contribution to the traces $T_1$, $T_2$, per vertex type $\tau$. The results in terms of the probability for finding a vertex with side parity $v = a$, $P$, are shown in Tab. I. Each row in the table represents the average incremental contribution to $T$ ($\equiv T_1 + T_2$) for a particular vertex, say $x$–a label for every row which includes the type of vertex, and whether it is occupied $a$ or $b$ (i.e. $x = $ "side $a, b$"). Then for each $x$, the coefficients for the terms involving the $1/6$ factor are the determined as increment in the value of $-\sum_i z_i/6$ (i.e. $T_1$) by computing the increase to $T_1$ adding an $x$. Specifically, we start with a "single vertex exclusion lattice" (a single inner vertex, its links, and neighboring vertices) which can be comprised of an $a$ or $b$ vertex (occurring with probabilities $P$ and $(1-P)$). For each case the increment in $T_1$ is computed after adding a second vertex $x$ to make $xa$ and $xb$ respectively. Each term enters with probability $P$ and $1-P$, thus giving the first two coefficients in the table (or a single number if the coefficients are the same). Similarly the coefficients for the second order terms are the increase of the sum $(-1/72)\sum_i z_i^2 + z_i(z_i - 1)$ (i.e. $T_2$) by comparing a two-vertex exclusion lattice comprised of $aa$, $ab$, $ba$, $bb$ (occurring with probabilities $P^2, P(1-P), P(1-P), (1-P)^2$), and adding a third vertex $x$ to make $xaa$, $xab$, $xba$, and $xbb$ respectively. When $x$ is a corner that is asymmetric, such as $\beta$ corners, an average (equal weight) over each side is also taken for each term.

Finally to obtain, $P$, we note that before the integration in Eq. A7, the Gaussian has mean $\mu \equiv PN_v$, which defines the mean-field probability $P$ used to self-consistently determine the $\tau$'s to be:

$$P = \frac{1}{2} + \frac{\tau_{side,a}(P) - \tau_{side,b}(P)}{4}. \quad (1)$$

With the estimates of Tab. I, numerically, $P$ is:

$$P = \frac{1}{4}\left(129 - \sqrt{16369}\right) \approx 0.2647. \quad (2)$$

## II. CONFIGURATIONS WHICH CANNOT BE REACHED BY THE RATIO METHOD UPDATES

To see that there are configurations which cannot be reached (for non-simply-connected regions) with our im-

TABLE I. Mean-field estimate of the average contribution to the trace $T$ per vertex of type $t$ with a value $a$ or $b$ respectively $\equiv \tau_{t,a(b)}$. $P$ is the likelihood of finding an $a$-valued vertex determined self-consistently (Eqs. 1 and 2). $\tau^*_{I\sigma,a}$ is computed for the vertex just before the corner. See text.

| | | |
|---|---|---|
| $\tau_{side,a}$ | $-\frac{6P+8(1-P)}{6}$ | $-\frac{34P^2+34P(1-P)+44(1-P)P+40(1-P)^2}{72}$ |
| $\tau_{side,b}$ | $-\frac{4}{6}$ | $-\frac{16P^2+16P(1-P)+12P(1-P)+12(1-P)^2}{72}$ |
| $\tau_{I\alpha,a}$ | $-\frac{8}{6}$ | $-\frac{48P^2+48P(1-P)+36P(1-P)+36(1-P)^2}{72}$ |
| $\tau_{I\alpha,b}$ | $-\frac{2P+4(1-P)}{6}$ | $-\frac{10P^2+10P(1-P)+20P(1-P)+16(1-P)^2}{72}$ |
| $\tau_{I\beta,a}$ | $-\frac{9P+10(1-P)}{6}$ | $-\frac{63P^2+71P(1-P)+34P(1-P)+48(1-P)^2}{72}$ |
| $\tau_{I\beta,b}$ | $-\frac{P+2(1-P)}{6}$ | $-\frac{1P^2+3P(1-P)+10P(1-P)+8(1-P)^2}{72}$ |
| $\tau^*_{I\sigma,a}$ | $-\frac{8}{6}$ | $-\frac{36P^2+36P(1-P)+48P(1-P)+48(1-P)^2}{72}$ |
| $\tau^*_{I\sigma,b}$ | $-\frac{4P+2(1-P)}{6}$ | $-\frac{24P^2+28P(1-P)+10P(1-P)+14(1-P)^2}{72}$ |
| $\tau_{II\alpha,a}$ | $-\frac{6}{6}$ | $-\frac{30P^2+34P(1-P)+22P(1-P)+22(1-P)^2}{72}$ |
| $\tau_{II\alpha,b}$ | $-\frac{4P+6(1-P)}{6}$ | $-\frac{20P^2+16P(1-P)+30P(1-P)+26(1-P)^2}{72}$ |
| $\tau_{II\beta,a}$ | $-\frac{7P+8(1-P)}{6}$ | $-\frac{41P^2+40P(1-P)+35P(1-P)+34(1-P)^2}{72}$ |
| $\tau_{II\beta,b}$ | $-\frac{3P+4(1-P)}{6}$ | $-\frac{13P^2+16P(1-P)+13P(1-P)+14(1-P)^2}{72}$ |
| $\tau_{II\sigma,a}$ | $-\frac{9P+10(1-P)}{6}$ | $-\frac{52P^2+52P(1-P)+62P(1-P)+58(1-P)^2}{72}$ |
| $\tau_{II\sigma,b}$ | $-\frac{2}{6}$ | $-\frac{6P^2+6P(1-P)+6P(1-P)+6(1-P)^2}{72}$ |

plementation of the ratio method, it is easiest to think of the transition loop picture between configurations.[1–3] Starting from a reference configuration, any other configuration may be described by flipping dimer occupations along closed loops (transition loops). Defining an operator $\sigma_{ij}$ which flips the dimer occupation on a link connecting $i$ to $j$, transition loops $\ell = \{i, j\}$ where $\{i, j\}$ are a set of links forming closed loops, can be found connecting any two configurations:

$$|C'\rangle = \prod_\ell \sigma_{ij}|C\rangle. \quad (3)$$

In this way, dimer coverings are seen as superpositions of loops over a reference configuration. Note that a plaquette flip can be considered a minimal (four-link) transition

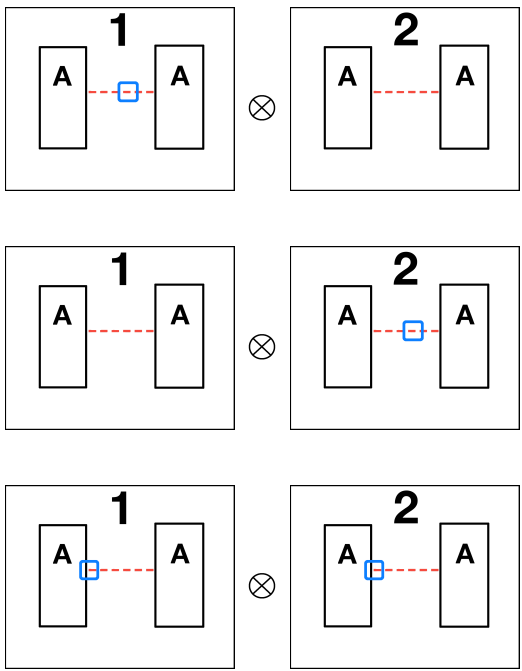

FIG. 1. (Color Online) A tensor product of copy 1 (left) and 2 (right), with transition loops for three types of Monte Carlo updates (top, middle, bottom) which affect the parity of transition-loop-crossings through the red dashed line.

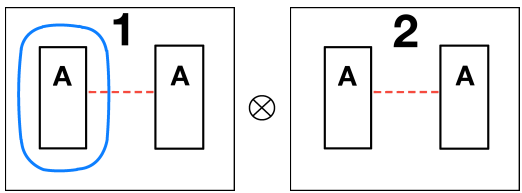

FIG. 2. (Color Online) A tensor product of copy 1 (left) and 2 (right) with a transition loop in blue which cannot be generated by the ratio method but which should belong to the (doubled) ground state. See text for explanation.

loop.

Now consider a region $A$ comprising two-disconnected parts, which constitutes a denominator or constrained region of the ratio method. Figure 1 shows illustrations of a doubled-dimer system ket, with such a disconnected region, and various plaquette flips as transition loops in blue. For such a bipartition, one can define a line in each copy connecting the two regions shown in the figure by the dashed red lines. The Monte Carlo updates that affect the number of transition-loop-crossings over these lines, is such that the sum of crossings over both copies can only change by two. To show this, Fig. 1 illustrates the three unique types of moves that could affect the crossings. In particular, because of the constrained plaquettes, the bottom panel includes plaquette flips that must be simultaneously applied to each copy.

However there do exist configurations which are both swappable but contain only an odd number of crossings. One example is shown in Fig. 2. Furthermore, these configurations could be reached if plaquette flips were unconstrained and therefore should be part of the ground state. Importantly there is no such conserved parity if the second disconnected piece of $A$ were not present. In that case any loop surrounding a region can be removed by outward plaquette flips and contracted to the identity in each copy *independently*.

These sorts of regions occur in the Levin-Wen construction,[4] and are therefore not amenable to our sampling via the ratio method without non-local updates.

* aselem@berkeley.edu
† Present Address: Department of Physics, University of Vermont, VT 05405, USA



\end{document}